\documentclass[preprint,12pt]{elsarticle}

\usepackage{amssymb}
\usepackage{amsmath}

\usepackage{natbib}
\usepackage{soul}
\usepackage{xcolor}

\usepackage{graphicx}
\usepackage{subcaption}
\usepackage{caption}
\usepackage{bm}
\usepackage[export]{adjustbox}  
\usepackage{makecell}
\usepackage{array}
\usepackage{tabularx}
\usepackage{comment}
\usepackage{algorithm}
\usepackage{algpseudocode}
\usepackage{placeins}

\usepackage{ulem}
\usepackage{url}
\journal{Computers \& Fluids}

\begin{document}

\begin{frontmatter}

\title{Adjoint-based shape optimization of a ship hull using a Conditional Variational Autoencoder (CVAE) assisted propulsion surrogate model}

\nonumnote{This is the accepted manuscript of an article accepted for publication in \textit{Computers \& Fluids}. \textcopyright~2026. This manuscript version is made available under the CC BY-NC-ND 4.0 license \url{https://creativecommons.org/licenses/by-nc-nd/4.0/}.}

\author[label1,label2]{Moloud Arian Maram\corref{cor1}}
\author[label2]{Georgios Bletsos\corref{cor2}}
\author[label1]{Thanh Tung Nguyen}
\author[label2]{Ahmed Hassan}
\author[label1]{Michael Palm}
\author[label2]{Thomas Rung}

\cortext[cor1]{Corresponding author: moloud.arianmaram@voith.com}
\cortext[cor2]{Corresponding author: george.bletsos@tuhh.de}
\nonumnote{The first two authors contributed equally to this work and share corresponding author responsibilities.}

\affiliation[label1]{organization={J.M. Voith SE \& Co. KG},
            addressline={Alexanderstrasse 2}, 
            city={Heidenheim},
            postcode={89522}, 
            state={Baden-Württemberg},
            country={Germany}} 

\affiliation[label2]{organization={Institute for Fluid Dynamics and Ship Theory, Hamburg University of Technology},
            addressline={Am Schwarzenberg‑Campus 4}, 
            city={Hamburg},
            postcode={21073}, 
            state={},
            country={Germany}}            

\begin{abstract}
Adjoint-based shape optimization of ship hulls is
 a powerful tool for addressing high-dimensional design problems in naval architecture, particularly in minimizing the ship resistance. However, its application to vessels that employ complex propulsion systems 
introduces significant challenges. They arise from the need for 
 transient simulations extending over long periods of time with small time steps and from the reverse temporal propagation  of the primal and adjoint  solutions.
These challenges place considerable demands on the required storage and computing power, which 
significantly hamper the use of adjoint methods in the industry.
To address this issue, we propose a machine learning-assisted optimization framework that employs a Conditional Variational Autoencoder-based surrogate model of the propulsion system. 
The surrogate model 
replicates the time-averaged flow field induced by a Voith Schneider Propeller and replaces the geometrically and time-resolved propeller with a data-driven approximation. 
Primal flow verification examples demonstrate that the surrogate model achieves significant computational savings while maintaining the necessary accuracy of the resolved propeller. %
Optimization studies demonstrate that neglecting the propulsion system can result in hull designs whose performance is inferior to that of the initial shape when subsequently validated using a numerically resolved propulsor.
In contrast, the proposed method produces shapes that actually achieve more than an 8\% reduction in resistance.

\end{abstract}


\end{frontmatter}

\section{Introduction}
 Shipping is one of the most popular modes of transportation, accounting for more than 80\% of global trade by volume \cite{1_1_unctad2023}. However, it also contributes significantly to greenhouse gases,   
NO$_x$ and SO$_x$ emissions, which pose severe environmental concerns \cite{1_2_smith2015, 1_3_bouman2017state}. Reducing the environmental footprint of shipping by increasing the hydrodynamic or propulsion efficiency can substantially reduce the fuel-related pollutant emissions of marine vessels \cite{1_4_LEE2014333}, and therefore has become a priority for shipbuilding. 
This has led to a growing interest in optimizations of
the hull shape and the  integration of the propeller in order to minimize the resistance and/or improve the propulsion efficiency, e.g. \cite{1_5_papanikolaou2014ship, 1_6_kim2012, stuck2011adjointII, springer2017simulation, 5_kuhl2022adjoint}. To this end,  adjoint-based methods have become essential due to their ability to compute gradients at a cost nearly independent of the number of design variables \cite{1_9_giles1997adjoint}. In marine applications, Springer and Urban proposed an adjoint-based framework for rigid body motion in multiphase flows, requiring only two static flow simulations \cite{1_springer2015adjoint}. Nazemian et al. achieved resistance reductions through adjoint solvers and mesh morphing \cite{2_nazemian2022shape, 3_nazemian2022shape_ship}, while Brenner et al. \cite{4_brenner2015parametric} integrated parametric modeling with CFD adjoint solvers to enable rapid sensitivity analysis. Kühl et al. advanced the research field by a consistent two-phase-flow method for free floating vessels, introducing a node-based adjoint method using the Steklov–Poincaré metric and improving solver stability on unstructured grids through elliptic relaxation \cite{5_kuhl2022adjoint, kuhl2021adjoint, 5_1_kuhl2025elliptic}. All these examples refer to steady-state optimizations and usually neglect propulsion influences.

\begin{figure}[h]
        \centering
        \includegraphics[width=11cm]{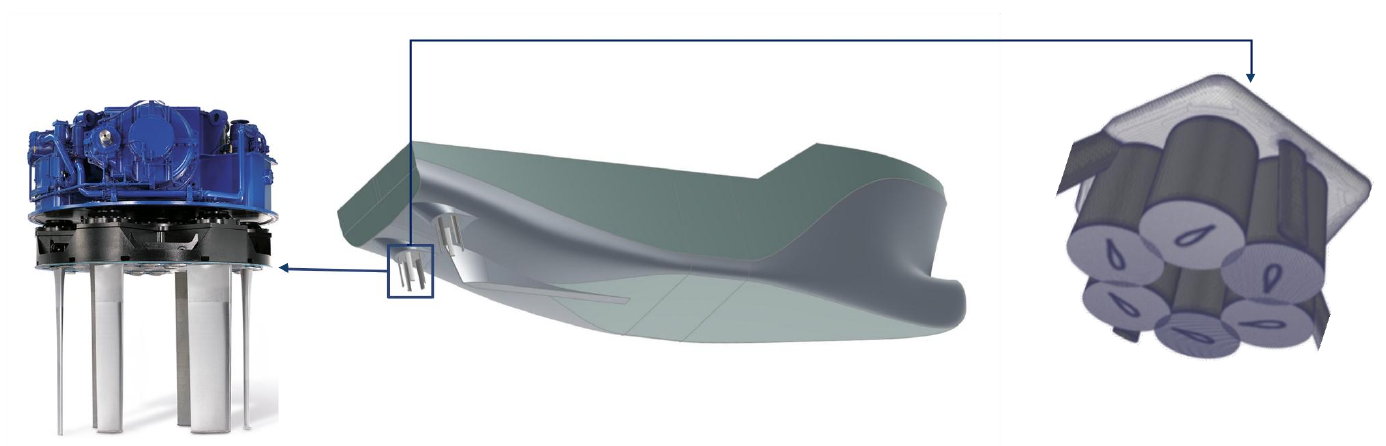}
        \caption{Geometry of the investigated Service Operation Vessel (SOV) equipped with two propeller units (center), a single Voith Schneider Propeller (VSP)  (left) and mesh components for a VSP unit with refinement near the propeller (right).}
        \label{fig:ship_vsp-a}
\end{figure}
    
The present study focuses on a marine vessel equipped with a Voith Schneider Propeller (VSP), cf. Fig. \ref{fig:ship_vsp-a}. The propulsion system can 
generate omnidirectional thrust through a cyclic blade motion \cite{5_4_voith2025marine, 5_5_jurgens2009voith}. Unlike conventional screw propellers, the VSP operates with blades rotating and simultaneously pitching in a vertical plane. The complex kinematics induces unsteady, spatially complex flow features, strong vortex shedding with pronounced blade–vortex interactions and pressure fluctuations 
\cite{5_6_voith2006propeller, 5_7_hu2022numerical, 5_8_esmailian2014numerical, 5_10_jurgens2009voith, 5_11_li2023numerical, 5_12_halder2018hydrodynamic}.
This presents two methodological challenges that adversely affect their 
numerical resolution in large-scale optimization applications:

\begin{itemize}
\item[1.]  
 A direct resolution of the rotating propeller would require unsteady adjoint simulations. These are difficult to perform due to the adjoint backward integration in time and the corresponding need to store the complete history of the primal flow field.

%
%
%
%
%
In case of large simulations with tens of millions of spatial grid points and hundreds of thousands of time steps, this is either unfeasible or yields an unacceptable reduction in computational efficiency for massively parallel applications on current high-performance computers. Proposed solutions are mostly based on check-pointing strategies, e.g. \cite{griewank2000algorithm, hinze2005revolve, Symes07, wang09}. These recompute a temporal segment of the primal solution starting from a nearby check-point and represent a compromise between computational and memory effort. However, the associated algorithmic effort and the considerable computational demands often limit their practical applicability.
An innovative alternative refers to singular value decomposition (SVD) methods, for constructing reduced-order models, e.g. 
 \cite{Sirovich87}. In the context of industrial adjoint shape optimizations, such lossy compression approaches must be implemented in a time-incrementing \cite{Brand02, Brand06, fareed2018incremental, fareed2019note, fareed2020error, vezyris2019incremental, Margetis23a} and parallel manner \cite{KuhlFisch23}.
\item[2.] Another critical aspect results from the broad spectrum of relevant time scales and the resulting large ratio of the total simulation duration to the time-step size. In naval engineering, this becomes particularly clear with the example of a free-floating ship: There, the time scale of the ship motion is several orders of magnitude larger than the time scale of the flow around the propeller. To reduce the associated effort, industrial CFD simulations often use multiple reference frames (MRFs) to simulate the flow around rotating components \cite{GOSMAN94}. 
Adjoint-based  studies using MRF methods to optimize wind turbines and ducted hydrokinetic devices are published in \cite{5_2_app8071112, 5_3_park2023cfd}.
Due to the perpendicular alignment of the cruise direction and the axis of rotation as well as the complex kinematics, the MRF concept is  not applicable to the VSP. An even simpler approach is the use of actuator-disc or body-force models  to represent  the propulsion. Such adjoint hull shape optimization approaches have been published in \cite{5_1_1_he2019design, stuck2011adjointII}. While actuator-disc surrogate models have proven effective for 
axial flows through   
screw-type rotors, they are not directly applicable to cycloidal systems such as the VSP. 
 
\end{itemize}
%

%
Within the framework of an adjoint analysis, the simulation of the resolved  or modeled VSP is therefore challenging. However, the influences of propulsion on the desired optimal shapes are of great importance. This can be seen, for example in Section~\ref{sec:opt_app}, where an optimization neglecting the propeller yields a hull shape that in fact results in an increased resistance when assessed under propulsion conditions. 
This typically also changes the load distribution across the hull surface, which --in turn-- can significantly influence both the floating position and the required propulsion power.

\subsection{Suggested Optimization Approach}
Due to the aforementioned challenges in resolving the VSP propeller within an adjoint-based optimization framework, we propose an AI-assisted methodology that integrates a machine learning (ML)-based surrogate model of the propulsor into a continuous adjoint-based shape optimization process. 

Conventional approaches either rely on fully resolving the rotating propeller geometry and the associated unsteady propeller--hull interaction, which requires computationally prohibitive unsteady adjoint simulations, or employ simplified models such as Multiple Reference Frame (MRF) or actuator-disk approaches. The latter significantly reduce computational cost but are not directly applicable to cycloidal propulsion systems like the VSP due to their complex kinematics and non-axial flow characteristics. 

In contrast, the proposed data-driven surrogate model is trained on time-averaged unsteady CFD data to emulate the propeller-induced flow field without requiring an explicit geometric representation of the propeller. This enables the inclusion of propulsion effects in adjoint-based optimization while avoiding both the cost of fully resolved unsteady simulations and the limitations of conventional simplified models.

 \smallskip
ML-based fluid dynamic applications have undergone an impressive development in the past. Recent studies demonstrate that ML-based models can reproduce turbulent flow fields with high accuracy while significantly reducing computational costs, see e.g. Vinuesa et al. \cite{6_vinuesa2024perspectives}, Duraisamy et al. \cite{7_1_duraisamy2019turbulence} or Brunton et al. \cite{7_2_brunton2020machine, brunton_aerospace}. Similarly, Thuerey et al. \cite{7_4_thuerey2021physics} demonstrated that generative networks can accurately predict complex 3D flow fields with minimal computational effort. In contrast to purely data-driven methods, physics-informed neural networks (PINNs) are frequently used to include physical constraints in the learning process \cite{RAISSI2019686, eivazi:pinn, liu2024multi}.

 \smallskip
The learning strategy employed in this work follows previously published data-driven methods for learning complex fluid dynamic fields using Convolutional Neural Networks (CNNs) \cite{agostini2020concept, swischuk2019projection, 10_1_pache2022data, kang2022physics}. However, in this work, machine-learning information is not used for a rapid, parameter-based  evaluation of designs, for example, in the context of onboard decision support systems \cite{10_1_pache2022data, Loft25}, but serves to support a flow simulation and adjoint analysis by 
adopting a Conditional Variational Autoencoder (CVAE) to learn the VSP-induced velocity field. In contrast to other autoencoder models, the CVAE features a latency space enriched with so-called labels, which are used to parameterize the fields to be learned. The enrichment replaces the additional fully connected neural networks that are used to map the parameters (labels) to the latent space suggested in other studies \cite{agostini2020concept, 10_1_pache2022data, kang2022physics, 9_3_schwarz2025machine}. In this work, we employ 11 labels, i.e. ten geometric parameters and one operating parameter that captures the cruising speed. 
Similar frameworks have been successfully applied in wind turbine condition monitoring, where CNN-based CVAEs effectively captured parameter–response relationships \cite{7_5_liu2022deep}. 
In the present investigation, the approach in \cite{7_5_liu2022deep} is extended by incorporating residual blocks and self-attention mechanisms 
 to enhance the accuracy of the output.

The transfer between the learned data and the optimization study is restricted to the flow velocities through the use of a meta-grid , i.e., a regular spatial grid used to represent the velocity field in the learning framework. The latter extends over a circular cylindrical domain that extends slightly beyond the area swept by the propeller.
By configuring the 11 data labels, the velocities are derived from the ML-based surrogate model at the discrete meta-mesh points. The information is subsequently interpolated to the CFD mesh used for the optimization studies, which does not resolve the propeller.
The introduction of ML-based velocities into the momentum equations uses an implicit enforcement. This not only enables a robust, tightly coupled treatment, which, in particular, eliminates the need to guarantee strictly continuity-compatible ML velocity fields, but 
also offers high flexibility in meshing. All other flow variables, i.e., pressure and turbulence variables, follow from their respective governing equations without any changes.
Building on this foundation, the model is tailored to efficiently handle %
a range of geometries and operating conditions, where the latter are represented by three distinct ship speeds,  
while preserving design sensitivity with sufficient accuracy.
Note that the training data is extracted from the immediate vicinity of the propeller and represent, among other things, the effective inflow to the propeller at a given ship speed. This inflow is induced by the ship’s geometry, which in turn is parameterized by the geometric parameters of the surrogate model.

The remainder of this paper is organized as follows: Section 2 details the methodology, including the development of the surrogate model, presenting the dataset generation, surrogate validation and integration to the CFD solver. The section closes with the detailed description of the employed adjoint-based optimization strategy. Section 3 describes the application of the proposed framework to a self-propelled hull optimization problem and performance assessment. Section 4 concludes with a summary of findings and discusses broader implications for CFD-based design in marine engineering.

\section{Autoencoder Surrogate Model}
\label{sec:method}

\begin{figure}[!h]
    \centering
    \includegraphics[width=0.5\linewidth]{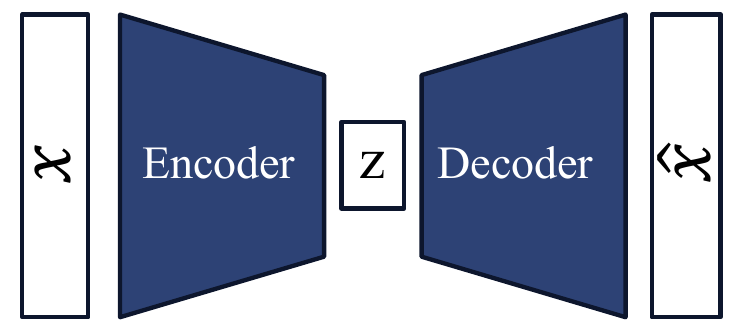}
    \caption{Employed basic autoencoder scheme: The input $X$ is compressed by the encoder network into a lower dimensional latent representation Z, and then reconstructed as $\hat{X}$ by the decoder network.}
    \label{fig:ae}
\end{figure}

The current study uses an autoencoder (AE), which is composed of an encoder and a decoder, as illustrated in Fig.~\ref{fig:ae}. The encoder maps the input $\mathbf{x}\in\mathbb{R}^N$ to a usually lower dimensional latent space representation $\mathbf{z}\in\mathbb{R}^m$ and the decoder subsequently maps  $\mathbf{z}$ to the full dimension to obtain a reconstruction of the input, $\hat{\mathbf{x}}\in\mathbb{R}^N$. Here $N$ is the total number of meta mesh grid points, and we consider 3 input vectors, which are the three Cartesian velocities. The encoder and decoder usually feature symmetric structures. 
All models in this work are based on a convolutional autoencoder (CAE), as the %
convolutional layers are usually better than fully connected layers at capturing spatial dependencies.

In particular, a Conditional Variational Autoencoder is employed as a surrogate model to approximate the velocity field in the vicinity of the VSP. 
Similar to the standard Variational Autoencoder (VAE), the CVAE is probabilistic and maps each input to a distribution parameterized by a mean and a variance, where the encoder performs stochastic sampling in the latent space and the decoder reconstructs or generates new flow fields from these samples~\cite{7_6_doersch2021tutorialvariationalautoencoders}. 
As illustrated in Fig.~\ref{fig:CVAE_structure}, the encoder  therefore provides two outputs, a mean $\boldsymbol{\mu}\in \mathbb{R}^m$ and a diagonal covariance matrix $\boldsymbol{\Sigma}\in\mathbb{R}^{m\times m}$, where the diagonal entries are $(\sigma_i)^2, i\in \{1,\dots,m\}$ and the $\log$-variance $\log (\boldsymbol{\sigma}^2)$ is utilized for stability reasons. 
The latent space representations are forced to follow
a probability distribution, which is assumed to be normally
distributed:  $\mathbf{z}\sim \mathcal{N}(\boldsymbol{\mu},\boldsymbol{\Sigma})$, 
using $\boldsymbol{\sigma} = \exp(0.5\cdot\log (\boldsymbol{\sigma}^2))$ and  generating $\boldsymbol{\epsilon} \sim \mathcal{N}(\mathbf{0},\mathbf{I})$, the computation of latent variables reads $\mathbf{z} = \boldsymbol{\mu} + \boldsymbol{\sigma} \odot \boldsymbol{\epsilon}$. %

Unlike the conventional VAE, however, the CVAE conditions both the encoder and the decoder on auxiliary inputs, enabling controlled generation of flow fields based on physical parameters.  
This conditioning mechanism is illustrated in the lower row of the encoder section in Fig.~\ref{fig:CVAE_structure},  and allows the surrogate model to learn a mapping from label–data pairs to flow-field distributions, thereby producing physically consistent outputs under new operating conditions. 

The training objective of the CVAE is defined through a composite loss function that balances two competing goals: reconstruction accuracy and latent-space regularization.  
The reconstruction loss is computed using the Mean Squared Error (MSE). It measures the difference between the predicted $\mathbf{\hat{x}}$ and reference flow fields $\mathbf{{x}}$, enforcing accurate reproduction of flow features, viz. 
\begin{equation}
\label{eq:MSE}
\mathrm{MSE} = \frac{1}{N} \sum_{i=1}^{N} (x_i - \hat{x}_i)^2 \, .
\end{equation}

A single MSE is computed over all cartesian velocity components (i.e. $x \in [$ $v_x$, $v_y$, $v_z]$) with equal weighting. We also explored component-wise weighted losses to account for differences in the magnitude of the velocities; however, no significant improvement was observed, and uniform weighting was therefore adopted.

To ensure a smooth and continuous latent representation, the Kullback–Leibler (KL) divergence term penalizes deviations of the learned latent distribution $q(\mathbf{z}|\mathbf{x})$ from the prior distribution $p(\mathbf{z}) = \mathcal{N}(0,\mathbf{I})$, i.e.

\begin{equation}
\label{eq:KL_final}
D_{KL}\big(q(\mathbf{z})\,||\,p(\mathbf{z})\big)
= \frac{\beta}{2}\Big[\mathrm{tr}(\bm{\Sigma}) + \bm{\mu}^\top\cdot\bm{\mu} - k - \log|(\bm{\Sigma})|\Big],
 \qquad \beta > 0 
\end{equation}
where $k$ is the dimensionality of the latent space. In the current study $k=2$ and $\beta = 1$. 
For $\beta= 1$, this is referred to as the VAE \cite{Kingma:2013} and $\beta\neq 1$ denotes the $\beta$-VAE~\cite{Higgins2016betaVAELB}.
The total loss function is  formulated as  
\begin{equation}
\label{eq:total_loss}
L = \mathrm{MSE} + D_{KL}\big(q(\mathbf{z})\,||\,p(\mathbf{z})\big),
\end{equation}
This composite loss ensures both an accurate flow-field reconstruction and a structured, continuous latent space suitable for interpolation and generalization.

\subsection{Configuration of the Conditional Variational Autoencoder}

The input to the surrogate model consists of the three cartesian velocity components 
$(v_x, v_y, v_z)$. Each component is normalized independently using the 
\textit{StandardScaler} method, which standardizes the data by subtracting 
the mean and dividing by the standard deviation. The same normalization 
strategy is also applied to the conditioning labels. To avoid data leakage, 
the normalization parameters (mean and standard deviation) are computed 
only from the training dataset and then consistently applied to the 
validation and test datasets. After this preprocessing step, the normalized 
velocity fields and labels are provided as inputs to the Conditional 
Variational Autoencoder model.

\begin{figure}
\centering
\includegraphics[width=1.4\linewidth, center]{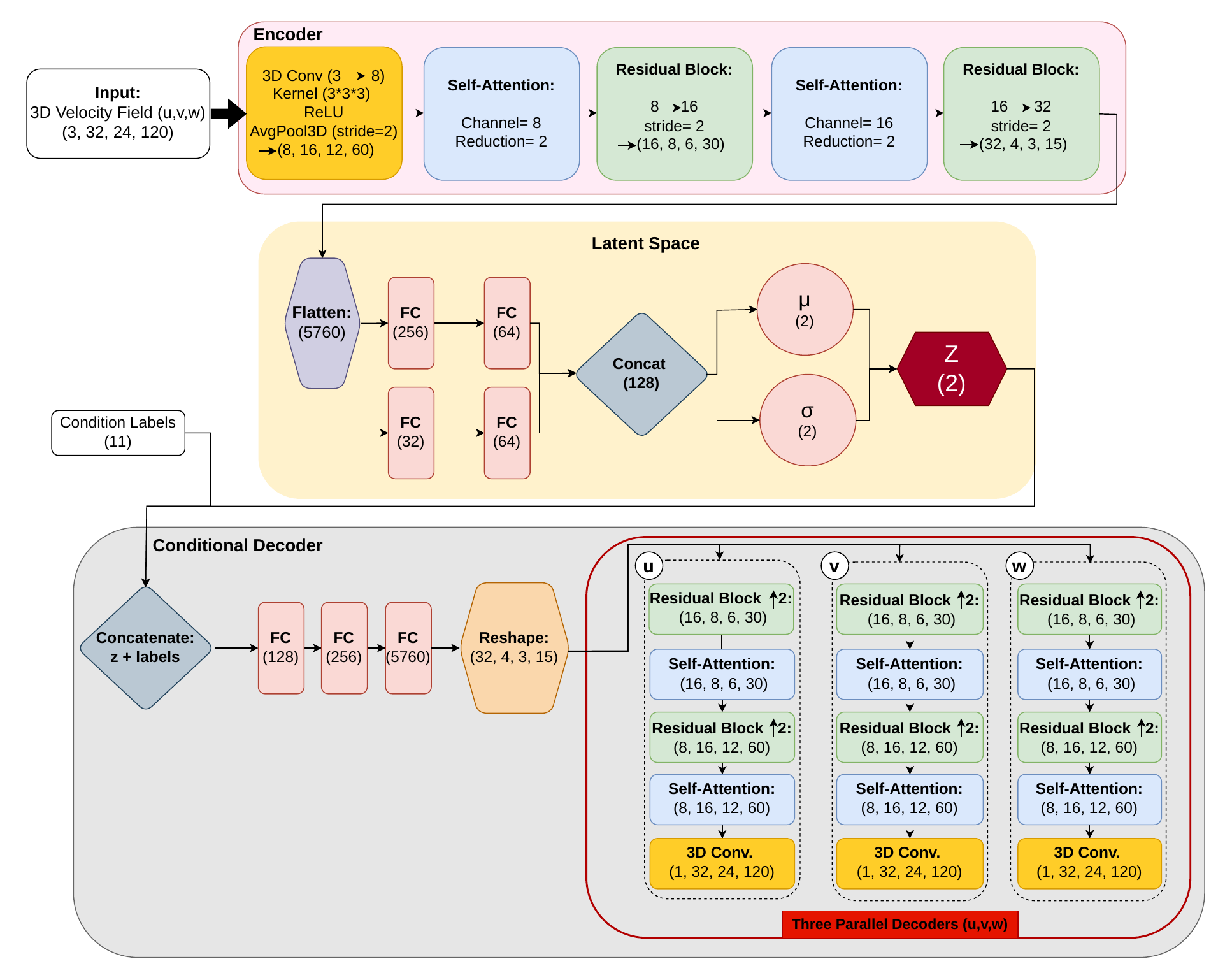}
\caption{ The CVAE architecture used in this study. The 3D velocity field is provided as input to the encoder and concatenated with 11 conditional labels. The latent representation $Z$ is reduced to two dimensions. The decoder reconstructs the 3D velocity field through three separate branches corresponding to the velocity components. Decoder residual blocks perform progressive spatial upsampling to recover the full resolution of the 3D field.}
\label{fig:CVAE_structure}
\end{figure}

Figure~\ref{fig:CVAE_structure} illustrates the architecture of the proposed Conditional Variational Autoencoder. 
The encoder processes a multi-channel 3D velocity input along with the conditioning vector, utilizing a combination of convolutional layers, residual blocks, single-head self-attention mechanisms, and fully connected layers. The conditioning vector is processed separately through fully connected layers and then appended to the encoded features. This combined representation is mapped to the latent space, producing the mean and standard deviation required for the reparameterization trick~\cite{Kingma:2013}, which enables differentiable stochastic training.

The latent vector is concatenated with the conditioning variables and subsequently expanded through fully connected layers to reconstruct the spatial feature representation.
The fully connected latent block operates on a compressed representation (dimension 5760) produced by the encoder and does not constitute a computational bottleneck, as the dominant computational cost arises from the 3D convolutional layers. Furthermore, the encoder progressively reduces the spatial resolution, allowing the latent dimensionality, and consequently the MLP size, to remain bounded when scaling to larger 3D domains.
The decoder mirrors the encoder architecture with slight modifications. In particular, the decoder residual blocks act as upsampling blocks, progressively increasing the spatial resolution of the feature maps to reconstruct the 3D field. Different decoder configurations were tested, including predicting the velocity components as a three-channel output. However, the adopted architecture showed better predictive accuracy and training stability, suggesting that the shared representation better captures the correlation between velocity components.
Since the three velocity components differ by several orders of magnitude, the conditioning variables are processed separately through three parallel branches, each composed of convolutional layers, self-attention modules, and residual blocks, dedicated to reconstructing one velocity component. Finally, the reconstructed velocity components are concatenated. 

\begin{figure} [htbp]
    \centering

    \begin{subfigure}[b]{0.7\linewidth} 
        \centering
        \includegraphics[width=\linewidth]{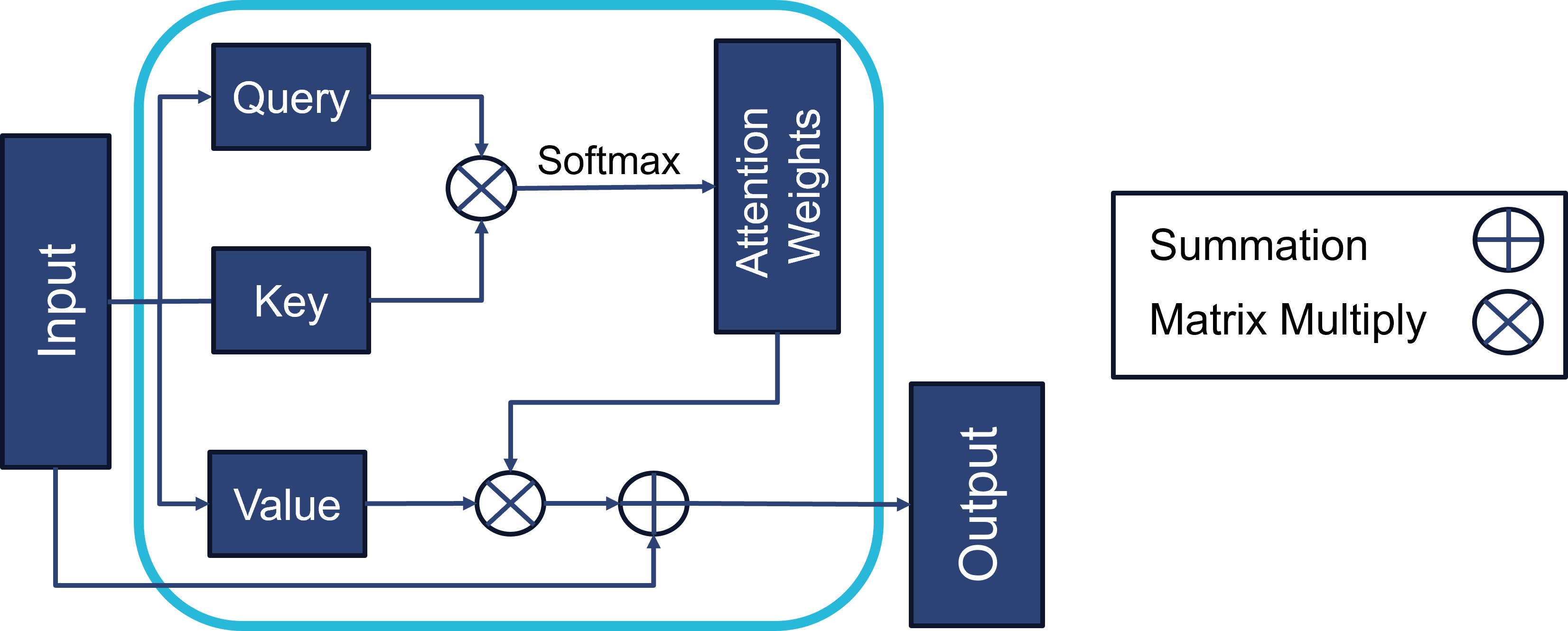}
        \caption{Self-attention mechanism}
        \label{fig:self_attention}
    \end{subfigure}

    \vspace{0.5cm} 

    \begin{subfigure}[b]{0.8\linewidth}
        \centering
        \includegraphics[width=\linewidth]{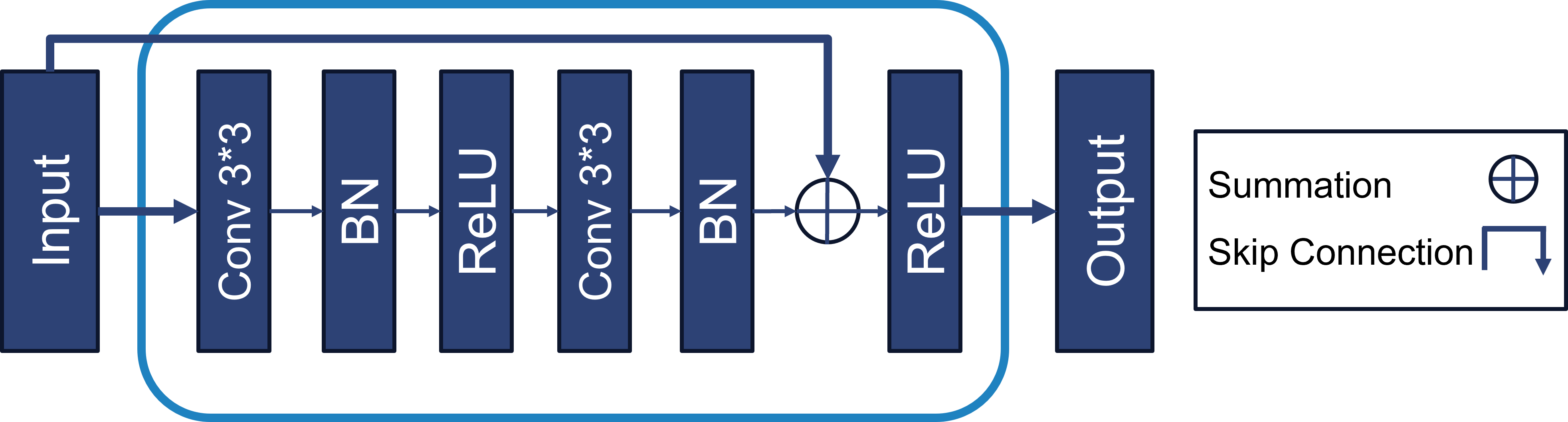}
        \caption{Basic residual block}
        \label{fig:basic_residual_block}
    \end{subfigure}
    
    \caption{Architecture of the model components. (top) Architecture of the self-attention mechanism, where the input is mapped to query, key, and value, and attention weights are computed using a softmax operation. The output is obtained by weighted summation. (bottom) basic residual block composed of convolution layers, batch normalization (BN), and ReLU activation, with a skip connection added to the output.}
    \label{fig:attention_vs_residual}
\end{figure}

Self-attention mechanisms, such as the non-local operation introduced by Wang et al.~\cite{8_2_1_Wang_2018_CVPR}, adaptively emphasize important spatial regions by weighting features based on global contextual relevance. Originally developed for natural language processing~\cite{9_ashish2017attention}, self-attention has been effectively applied in computational fluid dynamics to model non-local flow phenomena such as vortex shedding and flow separation. Peng et al.~\cite{9_1_peng2021attention} demonstrated its ability to capture long-range flow dependencies, Wu et al.~\cite{9_2_wu2021reduced} employed it within a convolutional autoencoder for improved flow reconstruction, and Schwarz et al.~\cite{9_3_schwarz2025machine} applied attention-based networks to predict fluid dynamic loads.
Figure~\ref{fig:self_attention} illustrates the self-attention mechanism. The query (Q) and key (K) matrices determine the relevance between different spatial locations through their dot-product similarity, while the value (V) matrix contains the corresponding feature information to be aggregated. This mechanism allows the network to selectively focus on important regions by weighting the value features according to the query–key attention scores.

Residual blocks facilitate the training of deeper networks by enabling stable gradient propagation \cite{10_he2016deep}. Schematic representations of the self-attention mechanism and residual block are provided in Fig~\ref{fig:basic_residual_block}. A typical block consists of two or more consecutive convolutional layers (Conv), each followed by batch normalization (BN) and a nonlinear activation function, such as ReLU in the present study.

Mind that in volumetric settings, the computational cost of self-attention increases rapidly with the spatial resolution.
 For a 3D feature map of size $D \times H \times W$, the number of spatial tokens is $N = D H W$, 
 and the complexity of the attention matrix scales as $\mathcal{O}(N^2)$.
  To reduce the associated computational cost, 
 the proposed module operates on spatially reduced feature maps. Specifically, the query ($Q$), key ($K$), and value ($V$) tensors are first down-sampled along each spatial dimension using an average pooling with a reduction factor of $r=2$
  before  the attention weights are computed. This reduces the number of spatial tokens by a factor of $r^3 = 8$, and the size of the attention matrix is reduced by approximately $r^6 = 64$ compared to full-resolution attention. In addition, the attention mechanism is applied only at intermediate feature resolutions within the encoder and decoder rather than on the original input grid. This approach significantly reduces the computational and memory overhead while enabling the network to capture long-range spatial dependencies in the flow field.

The proposed CNN-CVAE contains a total of approximately 
$3.15 \times 10^6$
trainable parameters. Based on the model summary, the estimated memory footprint is approximately $47.2$ MB for a batch size of 1, including input, activations, and parameters. The self-attention modules account for only a very small parameter overhead: the seven attention blocks together contain 2,760 trainable parameters, which corresponds to approximately 0.088\% of the entire model.

\subsection{Learning Rate} 
The learning rate is a key hyperparameter that controls the step size for updating model weights during training. To ensure both stability and efficient convergence, a 1-cycle learning rate schedule \cite{12_1_smith2019super} is generally employed over 1200 training epochs. The learning rate is initially set to $1 \cdot 10^{-4}$, peaks at $2 \cdot 10^{-4}$ by epoch 200, and gradually decreases to $5 \cdot 10^{-6}$ by the end of training. The model is optimized using the Adam algorithm \cite{14_kingma2014adam} with a mini-batch size of 4. Our preliminary investigations indicated that a latent dimension of two provided the best trade-off between reconstruction accuracy and model generalization. 

\section{CFD-based Methodology}
\label{cfd_dataset}
To construct the surrogate model, a CFD-based dataset was generated through transient numerical simulations. To limit the effort,
the current research is confined to single-phase flows around ships , where the vessel is fixed and no translational or rotational degrees of freedom are considered. The training, validation and test data refers to a sequence of offline simulations of the full-scale flow around a 92-meter-long Service Operation Vessel (SOV). 
The vessel is equipped with two five-bladed Voith Schneider propellers. Each VSP has a diameter of 2.65 meters and a blade length of 2.30 meters, and operates at a rotational speed of 79.28 rpm. The SOV cruises forward at 6.7-8.1 m/s, which yields a Reynolds number of {$Re=4.1-4.9 \cdot 10^8$} based on the length of the vessel and the cruising speed. Fluid properties corresponding to standard seawater conditions were assumed, and variations due to salinity were considered negligible  for this flow regime, as is customary in practice.
Since we only consider single-phase flows, the elevation of the free surface is neglected.

The employed CFD setup naturally limits the range of operating conditions and vessel configurations that can be modeled.
Obtained results are most relevant for Service Operation Vessels and similar offshore support vessels equipped with fully submerged Voith Schneider propellers, especially in twin-propeller configurations. 
The considered operating conditions are restricted to straight-ahead cruising at moderate advance ratios, i.e., the ratio between the cruising speed and the rotational speed of the propeller.
Because the analysis assumes a single-phase flow, its applicability is also limited to operating conditions in which the position of the free surface does not significantly affect the propeller flow.

\subsection{Governing Equations}
The offline simulations solve the three-dimensional, single-phase Reynolds-Averaged Navier–Stokes (RANS) equations, viz. 
\begin{align}
    R^\mathrm{p} &= \frac{\partial v_i}{\partial x_i} = 0 \label{eq:contrinuity} \\
    R_i^\mathrm{v} &= \rho\left[\frac{\partial v_i}{\partial t} + v_k \frac{\partial v_i}{\partial x_k}\right] + \frac{\partial}{\partial x_k}[p^{\mathrm{eff}} \delta_{ik} - 2\mu^\mathrm{eff}S_{ik}] = 0, \label{eq:momentum}
\end{align}
where, $v_i$ and $p^\mathrm{eff}$ correspond to the Reynolds-averaged velocity components and pressure, with the latter being augmented by the turbulent kinetic energy ($k$) term, $2\rho k/3$. The symbol $\delta_{ik}$ denotes the Kronecker delta. Based on the Boussinesq hypothesis, the effective viscosity $\mu^\mathrm{eff} = \mu + \mu^t$ of turbulent flows consists of the molecular viscosity and the turbulent contribution, which is identified by the solution of a two-equation turbulence model. The $k-\epsilon$ Boussinesq-viscosity model \cite{wilcox2006turbulence} is employed in all simulations conducted in this study. The components of the strain-rate tensor $S_{ik}$ are defined as

\begin{equation}
S_{ik} = \frac{1}{2}\left(
\frac{\partial v_i}{\partial x_k}
+ \frac{\partial v_k}{\partial x_i}
\right),
\label{eq:strainrate}
\end{equation}

Numerical results are obtained from a standard pressure-based, cell-centered finite volume method \cite{ferziger2020chapter13}. The approach is second-order accurate in space and time and capable of processing unstructured polyhedral meshes. The segregated procedure iterates the fields to convergence using a pressure-correction scheme.

\subsection{Computational Domain and Numerical Mesh}\label{comp_domain_num_mesh}
\begin{figure}[ht!]
    \centering
    \includegraphics[width=0.8\linewidth]{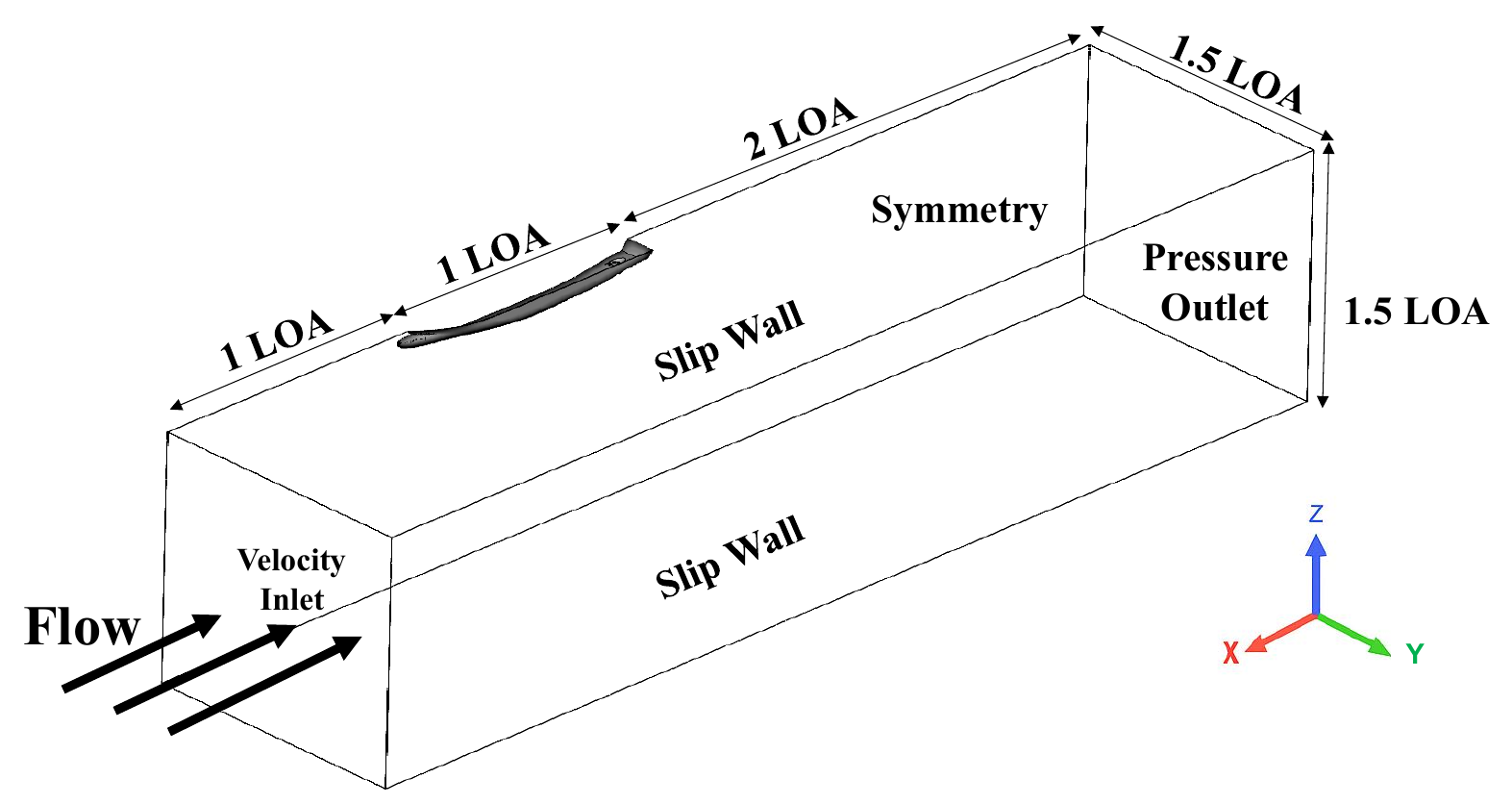}
    \caption{Computational domain and boundary conditions. The ship is located at distances of 1, 2, 1.5, and 1.5 LOA from the inlet, outlet, lateral side walls and bottom boundary, respectively. 
    }
    \label{fig:domain}
\end{figure}

The computational domain of the offline simulations includes a half-hull representation of 
 the submerged portion of the vessel, which is equipped with a Voith--Schneider propeller comprising five blades,
each with a blade length $L_{\mathrm{VSP}} = 2.3\,\mathrm{m}$
and a radius $r_{\mathrm{VSP}} = 1.33\,\mathrm{m}$. Boundary conditions are defined to reduce artificial reflections and maintain physical fidelity. A velocity inlet is located two ship lengths (LOA) upstream of the aft perpendicular, and a pressure outlet is placed two LOA downstream. A symmetry boundary condition is applied at the ship centerline. The lateral far-field boundary, located at 1.5 LOA from the centerline in the y-direction, is modeled as a slip wall. The hull surface is treated as a no-slip wall, and the top and bottom boundaries are defined as slip walls, with the bottom surface positioned 1.5 LOA below the waterline. The inflow is aligned with the negative x-axis. Figure~\ref{fig:domain} illustrates the computational domain and the VSP-equipped vessel.

\begin{figure}[ht!]
    \centering
    \includegraphics[width=1.0\linewidth]{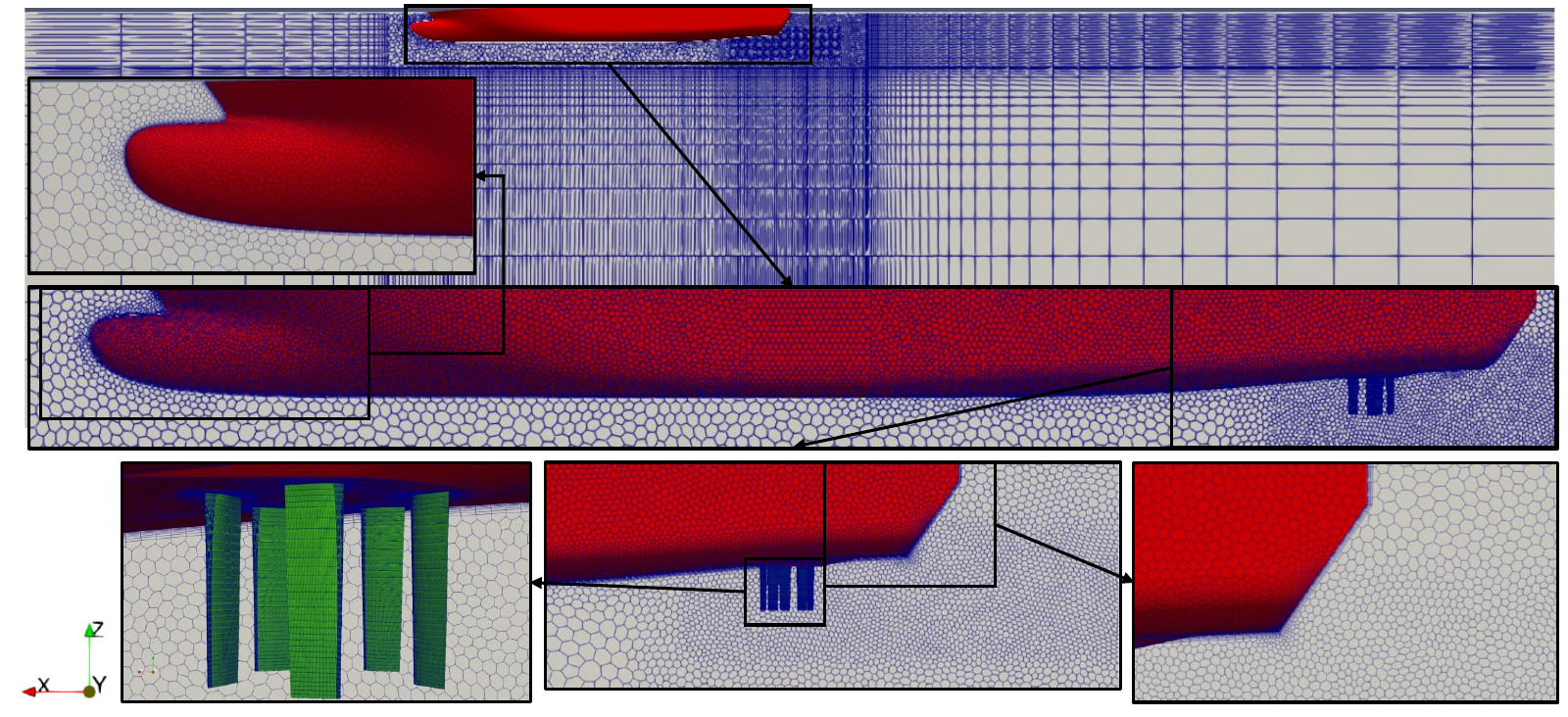}
    \caption{Computational mesh: the ship hull embedded in the unstructured polyhedral near-field mesh, full domain side view (top), with a close-up of the midship region (middle). Bottom row: close-up of the rotating VSP mesh cylinder containing the five blades (left), stern and VSP installation region (center), and bow region (right). The static unstructured polyhedral mesh ($\sim 0.85$ million cells) employs a base cell size of $1.0$ m, a hull surface cell size of $0.2$ m, and far-field cells up to $20$ m, with a prism-layer first-cell height of $y_1 = 5 \times 10^{-3}$ m. The rotating structured hexahedral mesh ($\sim 1.34$ million cells) consists of one outer cylinder enclosing five individual blades, with a prism-layer first-cell height of $y_1 = 1.5 \times 10^{-4}$ m.}
    \label{fig:mesh}
\end{figure}

 A sliding mesh technique is employed to resolve the VSP’s rotation in 1° increments per time step. To ensure accuracy and efficiency, the meshing strategy combines a structured hexahedral mesh around the VSP for capturing fine-scale flow features with an unstructured polyhedral mesh for the rest of the domain. The entire mesh consists of approximately 2.18 million cells and the propeller mesh consists of around 1.3 million cells, respectively. Wall functions are used to resolve the near wall flow, which corresponds to an approximate resolution of $y^+ \approx O$(100) along the walls. Figure~\ref{fig:mesh} illustrates the utilized mesh on the whole domain, the close-up images of the hull, bow, and propeller area.

\begin{figure}
    \centering
    \includegraphics[width=0.4\linewidth]{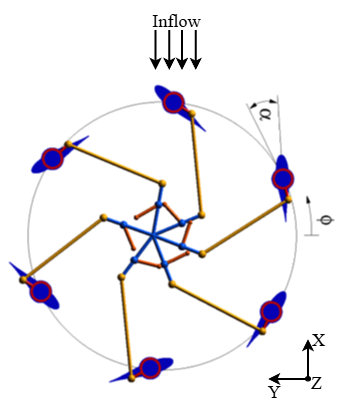}
    \caption{Schematic representation of the Voith Schneider Propeller (VSP) kinematics, illustrating the azimuthal blade position $\phi$ and the cyclic pitch angle $\alpha$. Each blade rotates around the propeller axis while oscillating about its own axis, resulting in a time-varying angle of attack.}
    \label{fig:vsp_rotation}
\end{figure}

The VSP blades undergo a cyclic pitch variation, where each blade oscillates about its own axis while simultaneously rotating around the propeller axis. This kinematic motion defines the instantaneous angle of attack of each blade and thereby determines the generated hydrodynamic forces and thrust direction. As illustrated in Fig.~\ref{fig:vsp_rotation}, the azimuthal position of a blade is denoted by the angle $\phi$, while the blade pitch angle $\alpha$ describes the instantaneous orientation of the blade relative to its trajectory. Within the sliding mesh CFD framework, this motion is implemented as a prescribed, time-dependent rotation of each blade about its own axis, synchronized with the overall propeller rotation.

To resolve the unsteady periodic flow, the simulation captures a periodic segment of 360°/5 blades = 72°, corresponding to one-fifth of a full revolution. This segment is discretized using a time-step resolution of 2°. After passing through the initial transient phase, the time-averaged flow field required for training is obtained by averaging the solution over these 36 steps, which effectively represents one complete revolution of the VSP.
It should be noted that, given the geometry in question, no additional periodic disturbances occur that would interfere with the blade-passage frequency. 
Therefore, the flow remains effectively periodic with respect to the blade motion, which justifies averaging over a blade cycle.

\subsection{Structure of Parameter Data}
\begin{figure}[ht!]
    \centering
    \includegraphics[width=1.0\linewidth]{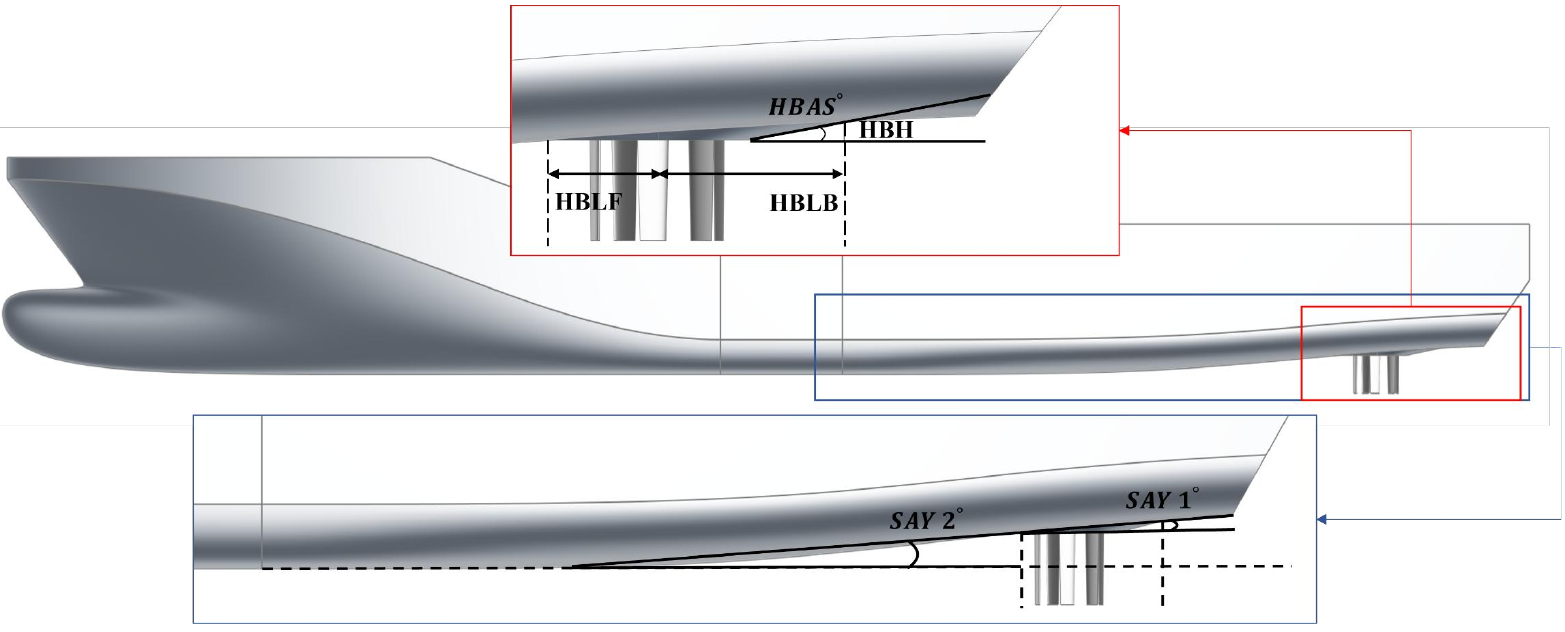}
    \caption{Illustration of the employed geometric parameters, including key aftship (SAY1, and SAY2) and headbox features (HBLF, HBLB, HBAS,and HBH), cf. Table \ref{tab:condition_ranges}.}
    \label{fig:geometric_parameters}
\end{figure}

To condition the surrogate model, multiple hull geometries, defined by variations of ten geometric parameters, are simulated under three forward speeds, i.e. 6.73, 7.42, 8.11 m/s. The geometric parameters are selected to capture both global and local features of the stern, cf. Fig.~\ref{fig:geometric_parameters}. The range of values for these parameters is summarized in Table~\ref{tab:condition_ranges}. They include the stern deadrise, stern inclination, stern angles Y1 and Y2 (SAY1, SAY2), and five variables related to the so called headbox, a structural extension above the propeller region that serves as a mounting point for the VSP. The latter consist of the front and back lengths (HBLF, HBLB), height (HBH), slope angle (HBAS), and transverse tilt angle (PhiX). HB\_Length is a non-dimensional parameter used to define the length of the headbox relative to the VSP diameter ($D_{\mathrm{VSP}}$). The actual headbox length is computed as HB\_Length~$\times$~$D_{\mathrm{VSP}}$.
The stern deadrise represents the transverse angle of the hull bottom at the stern, while the stern inclination defines the longitudinal slope of the aft hull section, both of which influence the wake and pressure field near the propeller. This parameterization enables a systematic variation of the aftship and flow-control features throughout the dataset.
 Note that three geometry parameters, namely the stern deadrise angle, the headbox angle (x), and the headbox length (front), remain frozen in the current study. 

\begin{table}[htbp!]
\centering
\caption{Range of parameters values considered by the data set.}
\begin{tabularx}{\linewidth}{l c c c}
\hline
\textbf{Parameter} & \textbf{Unit} & \textbf{Value Range} & \makecell{\textbf{Count of} \\ \textbf{Unique Values}} \\
\hline
Ship Speed (in $x$-direction)   & $m/s$    & 6.73--8.1 & 3.0 \\
Stern Deadrise                  & deg    & 0.0          & 1.0 \\
Stern Inclination               & deg    & 0.40--1.40   & 10.0 \\
Stern Angle Y1 (SAY1)           & deg    & 1.88--6.55   & 10.0 \\
Stern Angle Y2 (SAY2)           & deg    & 5.29--17.95  & 10.0 \\
HB\_Length                      & -      & 1.0--2.0     & 6.0 \\
Headbox Angle X (PhiX)          & deg    & 0.0          & 1.0 \\
Headbox Angle Slope (HBAS)      & deg    & 6.91--35.38  & 26.0 \\
Headbox Height (HBH)            & m      & 0.42--1.88   & 29.0 \\
Headbox Length Back (HBLB)      & m      & 3.81--6.43   & 18.0 \\
Headbox Length Front (HBLF)     & m      & 2.12         & 1.0 \\
\hline
\end{tabularx}
\label{tab:condition_ranges}
\end{table}

For each simulation, the quality of the solution is ensured based on the convergence of the predicted hull force and propeller thrust.

\subsection{Meta-Grid}
\begin{figure}
    \centering
    \includegraphics[width=0.9\linewidth]{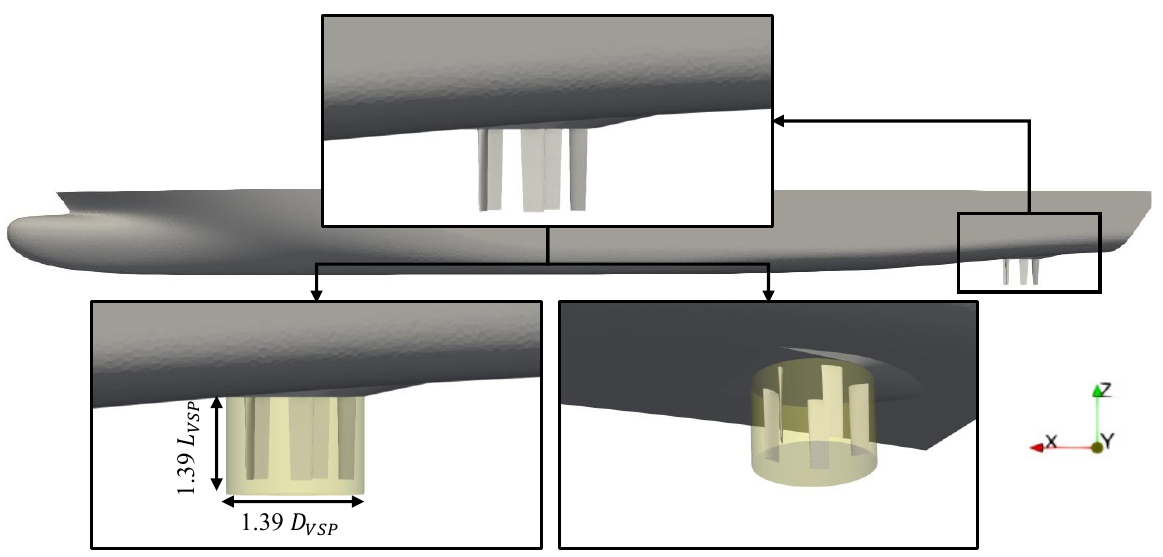}
    \caption{Configuration of the meta-grid around the VSP. The meta-grid radius and height are set to 1.39 times the propeller diameter and height, respectively.}
    \label{fig:vsp_cylinder}
\end{figure}

To prepare the communication between the CFD procedures and the surrogate model for the employed machine-learning strategy, a consistent and structured data representation is required. The latter involves a meta-grid which is used to connect the offline CFD-procedures during the training of propulsion influences and the online CFD-procedures during the optimization.
To this end, a cylindrical control domain is defined around the VSP. 
The coordinate system is oriented along the rotational axis of the propeller, 
with the origin located at the propeller mounting point and the positive axial 
direction pointing from the blade root toward the blade tip. 
The domain dimensions are given by 
$z = 1.39 L_{\mathrm{VSP}}$ in the axial direction and 
$r = 1.39 r_{\mathrm{VSP}}$ in the radial direction, 
capturing flow field variations beyond the blade tip and blade orbit, 
particularly at the upper axial limit ($z = 3.1\,\mathrm{m}$).
Figure~\ref{fig:vsp_cylinder}  depicts the extension of the meta-grid and its positioning relative to the VSP. Mind that the spacings in the meta-grid are deliberately chosen non-equidistant to provide higher resolution in regions where the velocity gradients are larger, particularly in the $\theta$-direction.
{\color{purple} }

\subsubsection{Integration of offline CFD results into surrogate model}
\begin{figure}
    \centering
    \includegraphics[width=0.5\linewidth]{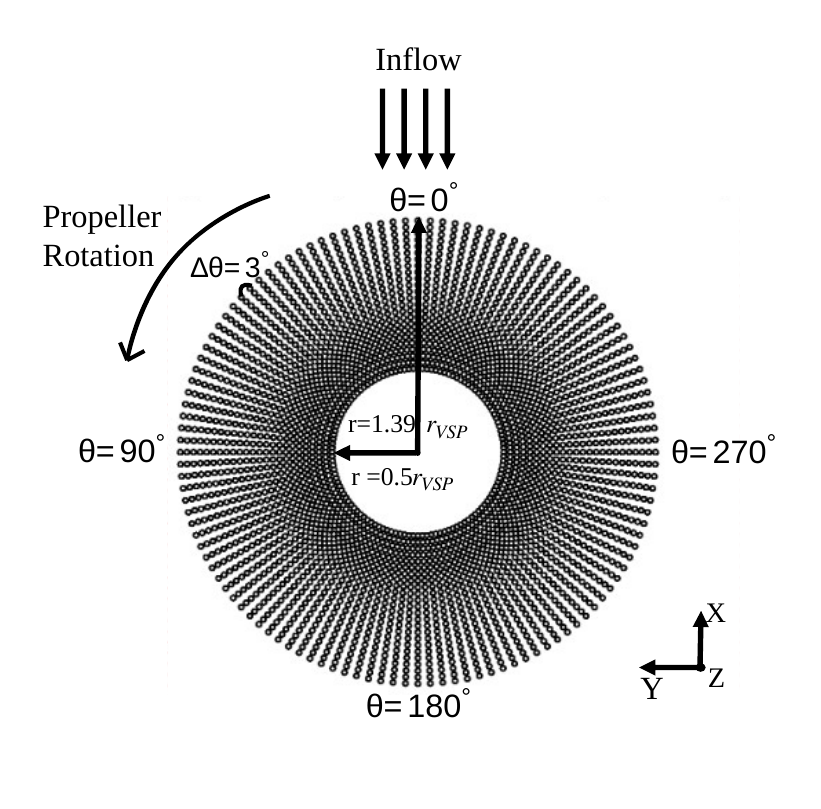}
    \caption{2D sketch of the meta-grid spacing and angular discretization. The angular coordinate is defined such that $\phi = 0^\circ$ corresponds to the forward (bow) direction and $\phi = 180^\circ$ to the aft (stern) direction. The propeller rotates counterclockwise.}
    \label{fig:meta_grid_2D}
\end{figure}

The extraction nodes in the CFD simulation are fixed and defined based on the meta mesh which is overlaid on the computational domain. For each time step, the unstructured CFD data are mapped to the structured meta-grid during training using the query functionality in Ensight, with default settings.
Figure~\ref{fig:meta_grid_2D} illustrates a sketch of the spacing of the structured meta-grid in a two-dimensional plane. This preserves key flow features while maintaining consistent formatting across all samples.  Depending on the blade rotation of the VSP, some nodes at certain time steps lie inside the blade region and therefore are not part of the computational domain. 
These nodes are excluded for that specific time step during the time-averaging process, where the averaging interval corresponds to one blade period of 72°, with an angular increment of 2°, resulting in a total of 36 time steps.
During the time averaging, only the valid nodes (where no blade was present) are taken into account. 
%

%
%
%

%

\subsubsection{Integration of surrogate model into CFD software}
The surrogate model should facilitate the simulation-based optimization of hull shapes by seamlessly integrating non-resolved propulsion influences. 
It outputs the 3D velocity field, given the parameters described in Table~\ref{tab:condition_ranges}, in a structured cylindrical meta-grid. 
To allow for generalization, we consider now a new CFD mesh, outside of the pool of meshes employed for the generation of training, validation, and test data, which discretizes the computational domain without resolving the propeller.
Based on the position of the headbox in the computational domain and its inclination w.r.t. the main flow direction, the meta-grid is translated and rotated to the position where the propeller would have been mounted on. Once this transformation is executed, the velocities are mapped from the meta-grid to the new CFD mesh using 
a nearest-neighbor interpolation, i.e.,
\begin{equation}
    \Tilde{\phi}(\bm{x}) = \phi(\Tilde{\bm{x}}_j), \quad \text{where} \quad j=\arg \min_i |\bm{x} - \Tilde{\bm{x}}_i| \label{eq:nearest_neigh}
\end{equation}
with $\Tilde{\phi}(\bm{x})$ denoting the interpolated quantity, here, one of the 3 Cartesian velocity components, at a cell-center $\bm{x}$ of the CFD mesh and $\phi(\Tilde{\bm{x}})$ the surrogate model predicted quantity on a cell-center $\Tilde{\bm{x}}$ of the meta-grid. 
The interpolation strategy is only realized for cells in the CFD mesh that lie within the limits of the rotated and translated meta-grid. 

Superficially, the integration of the surrogate model into the CFD domain aims to manipulate the solution $\phi$ at the prescribed location, so that it matches the interpolated output $\Tilde{\phi}$. A direct approach would simply overwrite the solution at the interpolated cells with $\Tilde{\phi}$. However, this approach is conceptually flawed because of the strong coupling between the velocity and the other field variables, in particular the pressure. 
Instead, we apply the approach of solution forcing \cite{ferziger2020chapter13}. In this approach the solution of the RANS equations ($\phi$) is naturally blended with the interpolated quantity ($\Tilde{\phi}$) by adding a source term to the discretized momentum equations of the following form
\begin{equation}
    A_P\phi_P + \sum_{N(P)} A_N \phi_N = Q_P + \alpha \left[\rho V (\Tilde{\phi} - \phi)\right]_P,\label{eq:forcing_orig}
\end{equation}
where $\alpha$ stands for a forcing coefficient and $V$ for the volume of a computational cell with centroid $P$. $A_P$, $A_N$ and $Q_P$ refer to the main diagonal coefficient, neighbor coefficient and source term, respectively, arising from the finite volume discretization of the momentum equations. It is easy to see, that $\alpha$ is not dimensionless, thus making its selection not only problem- but also location- and time- dependent. To avoid this, we modify the forcing in an implicit manner by setting 
\begin{align}
    A_P\phi_P + \sum_{N(P)} A_N \phi_N &= Q_P + \hat{\alpha} A_P \left[(\Tilde{\phi} - \phi)\right]_P \nonumber \\ 
    A_P(1+\hat{\alpha})\phi_P + \sum_{N(P)} A_N \phi_N &=Q_P + \hat{\alpha}A_P\Tilde{\phi}_P,\label{eq:forcing_mod}
\end{align}
where $\hat{\alpha} \geq 0$ is a dimensionless forcing coefficient. Note  $\hat{\alpha} = 0$ is imposed in the cells outside of the limits of the meta-grid. Due to the proportionality of the forcing to the main diagonal coefficient, the implicit forcing approach
(\ref{eq:forcing_mod}) is space (and time) dependent, in contrast to the explicit approach (\ref{eq:forcing_orig}). Moreover, the implicit approach strengthens the main diagonal and thus enhances the robustness of the iterative procedure.
Strictly speaking, the condition $\phi = \tilde \phi$ can only be satisfied for (infinitely) large values of $\alpha$ or $\hat{\alpha}$. However, this is generally undesirable and would indeed be useless, since neither the surrogate model nor the interpolations used are capable of generating divergence-free velocities that are compatible to the discretized continuity equation (\ref{eq:contrinuity}) in the target mesh. The use of PINN  strategies to obtain divergence-free meta-grid velocity fields is conceivable, but not very helpful, since even marginal deviations from the divergence-free state can prevent the convergence of the numerical method.
Note that 
the  dimensionless property $\hat{\alpha}$ is deliberately assigned to values of the order $O(10^{-1})$ which provides room for deviations from the target field $\tilde \phi$ that are required to preserve the continuity condition (\ref{eq:contrinuity}). Based on our experiments, cf. Section~\ref{sec:verification_SM_CFD}, a forcing parameter in the range of $\hat{\alpha} \in [0.1, 1]$ is sufficient to provide the desirable forcing behavior.

Boundary cells lying within the limits of the meta-grid refer to the main rotor of the VSP. In these cells Dirichlet boundary conditions are explicitly applied for velocity based on the rotational speed of the propeller.

%

\section{Verification and Validation}

To assess the capabilities of 
the surrogate model, a structured training and evaluation procedure was employed. For this purpose, the data set consisting of 180 samples was divided into training (80\%), validation (10\%), and test subsets (10\%).
The model was trained using the training set, incorporating both flow field data and the corresponding conditioning labels. A companion validation was conducted every 10 epochs to monitor convergence and mitigate overfitting. After the training, the performance of the model was evaluated using the previously unseen input conditions of the test subset.

\subsection{Validation of the Surrogate Model}

To evaluate the performance of the proposed CNN-CVAE model, several error metrics were calculated for the entire test dataset. To this end, the predicted velocity fields were compared with the corresponding CFD reference solutions. The evaluation includes the individual Cartesian velocity components $(v_x, v_y, v_z)$, the velocity magnitude, and the velocity divergence. The analysis is based on conventional L$_2$ and L$_\infty$ error norms in absolute and relative form. 
The obtained results indicate that the proposed CNN-CVAE model achieves good agreement with the CFD reference data across all velocity components. The relative $L_2$ errors remain below $4\%$ for $v_x$, $v_y$, and $v_z$, while the velocity magnitude error is limited to $1.24\%$, demonstrating accurate reconstruction of the overall flow field. Although the divergence-related errors are comparatively higher, the results still confirm that the model preserves the main physical characteristics of the flow with reasonable accuracy.

\begin{table}[h]
\centering
\caption{Quantitative error metrics computed over the full test dataset.}
\label{tab:test_metrics}
\begin{tabular}{lccc}
\hline
& abs. $L_2$ & rel. $L_2$ & abs. $L_\infty$ \\
\hline
$v_x$ & $1.05 \times 10^{-1}$ & $1.31\%$ & \\
$v_y$ & $5.46 \times 10^{-2}$ & $3.74\%$ & \\
$v_z$ & $3.20 \times 10^{-2}$ & $2.92\%$ & \\[2mm]
Velocity Divergence & $6.69 \times 10^{-2}$ & $6.43\%$ & $1.16$ \\
Velocity Magnitude  &                       &   $1.24\%$    & \\
\hline
\end{tabular}
\end{table}

In addition to the global error metrics, a statistical analysis of the prediction errors was performed across the entire test dataset. For each test case, the prediction error was first computed at every spatial grid point of the 3D velocity field. Subsequently, spatial statistics were evaluated over the computational domain, including the mean, median, and 95th percentile of the absolute error. These spatial statistics were then aggregated across all test samples to characterize the distribution of prediction errors across the dataset.
Table~\ref{tab:error_distribution} summarizes the distribution of per-sample spatial mean absolute errors across the test set. 
Overall, the axial velocity component $v_x$ exhibits the largest prediction error, which is expected as it corresponds to the dominant 
streamwise flow induced by the propeller. In contrast, the transverse velocity components $v_y$ and $v_z$ show significantly smaller 
prediction errors. The divergence error remains relatively low across the dataset, indicating that the predicted velocity fields 
approximately preserve the incompressibility condition.

\begin{table}[t]
\centering
\caption{Distribution of per-sample spatial mean absolute errors across the test set. 
For each test sample, the mean absolute error (MAE) was computed over the 3D field, 
and the distribution of these errors across all test samples is summarized using the mean, median, and 95th percentile.}
\label{tab:error_distribution}
\begin{tabular}{lccc}
\hline
\textbf{Quantity} & \textbf{Mean} & \textbf{Median} & \textbf{95th Percentile} \\
\hline
$v_x$ & $7.63 \times 10^{-2}$ & $5.49 \times 10^{-2}$ & $8.99 \times 10^{-2}$ \\
$v_y$ & $3.73 \times 10^{-2}$ & $3.37 \times 10^{-2}$ & $4.25 \times 10^{-2}$ \\
$v_z$ & $2.12 \times 10^{-2}$ & $1.89 \times 10^{-2}$ & $2.56 \times 10^{-2}$ \\
$|\mathbf{v}|$ & $7.45 \times 10^{-2}$ & $5.37 \times 10^{-2}$ & $8.75 \times 10^{-2}$ \\
Divergence & $4.53 \times 10^{-2}$ & $4.29 \times 10^{-2}$ & $5.49 \times 10^{-2}$ \\
\hline
\end{tabular}
\end{table}

\begin{figure}  
    \centering
    \includegraphics[width=1.2\linewidth,center]{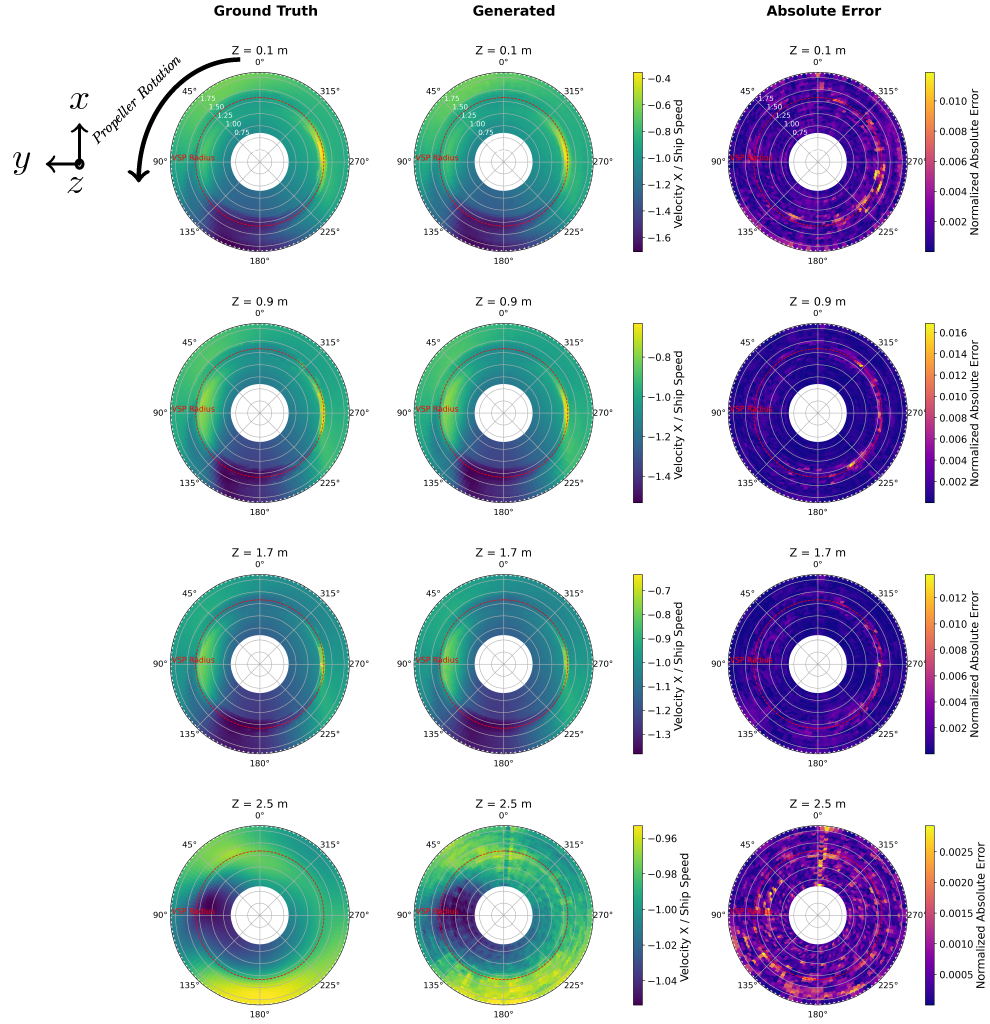}
    \caption{Comparison of  $x$-velocity contours obtained from the time-averaged CFD (left) and the surrogate model (center) normalized with the ship speed in addition to the corresponding normalized error magnitude contours (right) for the test sample in the meta-grid.}
    \label{fig:v_x}
\end{figure}

\begin{figure}
    \centering
     \includegraphics[width=1.2\linewidth,center]{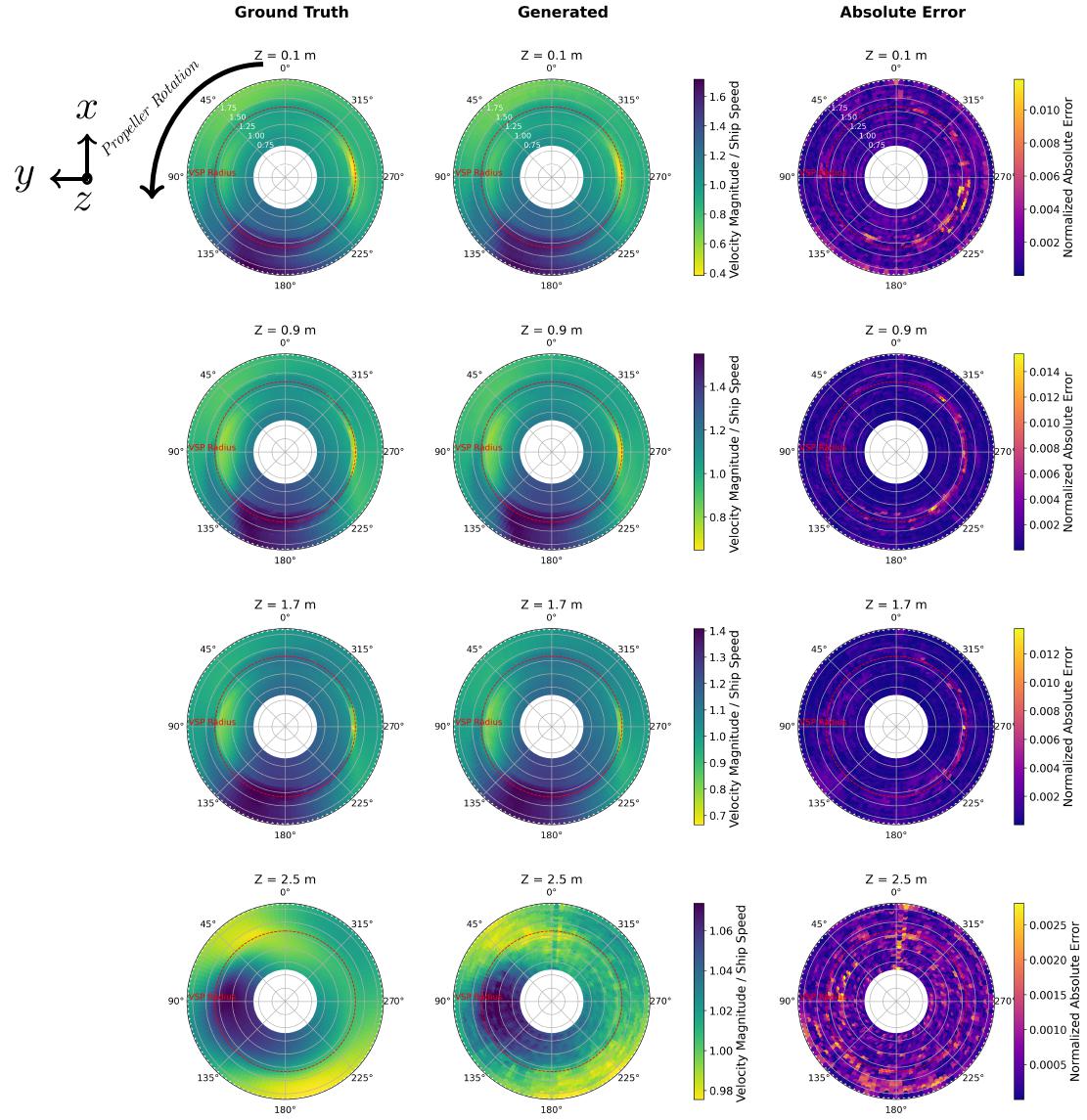}
    \caption{Comparison of velocity magnitude contours obtained from the time-averaged CFD (left) and the surrogate model (center) normalized with the ship speed in addition to the corresponding normalized error magnitude contours (right) for the test sample in the meta-grid.}
    \label{fig:v_mag}
\end{figure}

The results are presented for one representative sample of the test subset.
The corresponding %
parameters for this test sample are summarized in Table~\ref{tab:condition_values}.
\begin{table}[ht]
\centering
\caption{Conditioning parameters used for the selected test sample.}
\label{tab:condition_values}
\begin{tabular}{lll}
\hline
\textbf{Parameter} & \textbf{Unit} & \textbf{Value} \\
\hline
Ship Speed ($x$-direction)        & m/s     & 7.42   \\
Stern Deadrise     & deg    & 0.0   \\
Stern Inclination  & deg    & 1.067   \\
SAY1               & deg    & 4.99   \\
SAY2               & deg    & 13.87   \\
HB\_Length         & -      & 1.2   \\
PhiX               & deg    & 0.0   \\
HBAS               & deg    & 24.88   \\
HBH                & m      & 1.2   \\
HBLB               & m      & 4.42   \\
HBLF               & m      & 2.12   \\
\hline
\end{tabular}
\end{table}
The analysis focuses on the axial velocity component, the velocity magnitude (both normalized by the ship speed), and the divergence of velocity
 in four axial slices, i.e., $z=0.1, 0.9, 1.7$ and $2.5$ m below the mounting point of the VSP, within the cylindrical meta-grid domain.
Figures~\ref{fig:v_x} and~\ref{fig:v_mag} compare the results of the offline CFD simulations interpolated to the meta-grid (left) with results obtained from the surrogate model (center) and also outline the respective normalized error magnitudes (right). The velocity
error is normalized by the ship speed to obtain a relative error metric; this
normalization does not imply an increase of the absolute error with increasing
ship speed.
 The contour plots are presented in the local cylindrical coordinate system aligned with the propeller rotational axis. The axial direction of this system (z-axis) is aligned with the global streamwise direction (x-axis), although the coordinate origins differ. In this representation, the upstream direction corresponds to the top (0°) and the downstream direction to the bottom (180°). The 
 flow enters the cylindrical domain around the observed, counter-clockwise rotating, port propeller from above (0°) and the accelerated flow leaves the domain at the bottom with a slight inclination away from the symmetry plane, i.e., approximately at the 160° location. The dashed red circular line corresponds to the line around which the pitching blades rotate. Mind that the bottom graphs refer to a plane underneath the propeller that is not swept by the propeller blades, which only reach to $z=2.3$ m.

The axial velocity $v_x$ dominates the forward-directed propulsion and also exhibits the largest variations in the domain covered by the meta-grid. Accurately capturing this component is critical for characterizing the thrust and the wake behavior.

 The results displayed demonstrate that the surrogate model reconstructs both global and local flow patterns with high fidelity. The model mimics the generation of axial momentum by the propeller quite well and also captures the areas of strong deceleration near the blade paths. The errors remain locally confined and are typically at a low level of approximately 1–2\% of the ship speed. Local maximum errors occur in areas with large spatial gradients and are approximately 10\% of the ship's speed. The error magnitude generally increases 
 closer to the headbox.

\medskip
Another approach to assess the predicted velocity components is to study the velocity divergence, i.e., 
how well the  velocity field predicted by the surrogate model satisfies the conservation of volume in comparison to the time-averaged interpolated CFD results.
Strictly speaking, the unsteady CFD results are conservative, but the time-averaged interpolated CFD results are not necessarily conservative, in particular, if one considers the fluid homogenization of the regions temporarily swept by the propeller blades.

\begin{figure}
    \centering
    \hspace*{-1.3 cm}
    \includegraphics[width=1.2\linewidth]{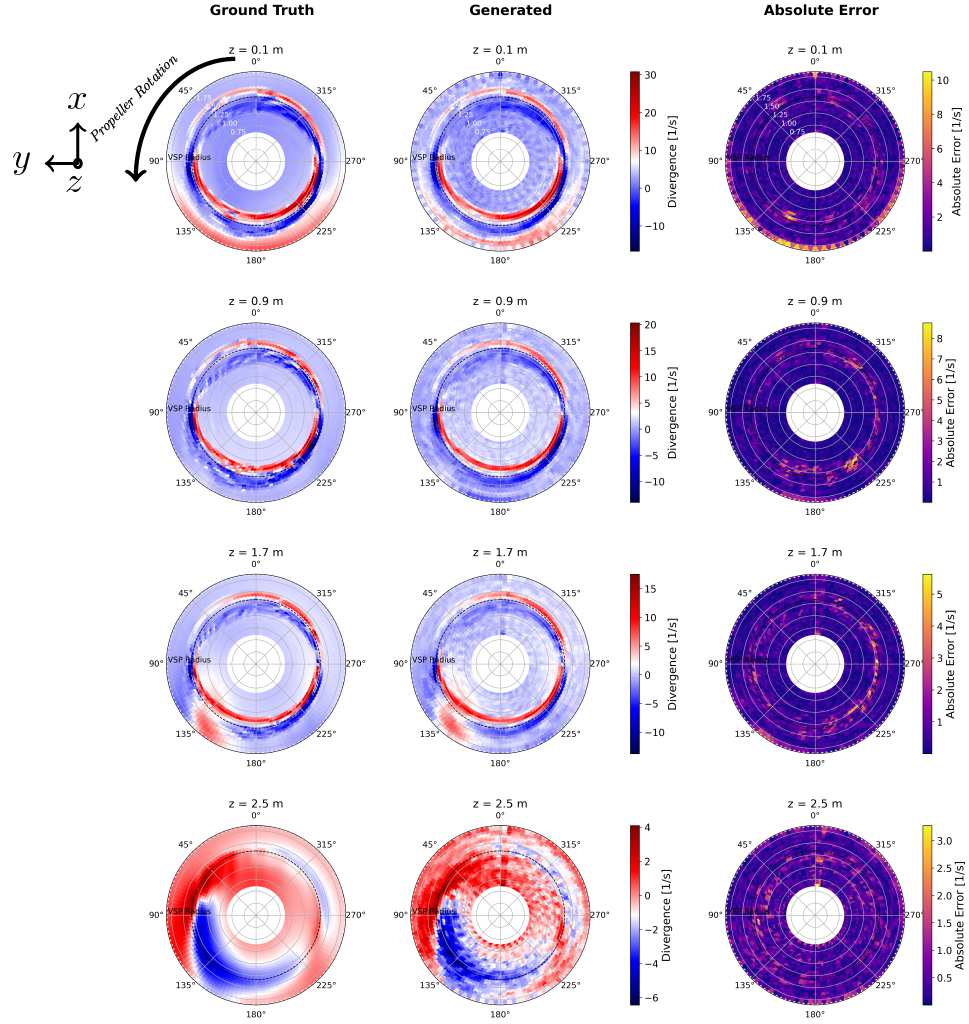}
    \caption{ Comparison of velocity divergence contours obtained by 2nd-order Finite Differences from the time-averaged CFD (left) and the surrogate model results (center)  with corresponding error magnitude contours (right) for a test sample in the meta-grid.
}
    \label{fig:velocity_Divergence}
\end{figure}

For this purpose, the velocity divergence in the meta-grid was calculated twice, once for the time-averaged CFD data to be learned and once for the predictions of the surrogate model. 
The resulting contour plots of the velocity divergence and the corresponding error magnitudes for selected $z$-planes are presented in Fig.~\ref{fig:velocity_Divergence}, whose structure is similar to that of Figs.~\ref{fig:v_x} and \ref{fig:v_mag}.

In general, significant non-zero divergence levels are observed.
The largest levels are located near the circumference of the rotor (cf. dashed line),
while the outer and inner radial zones remain mostly divergence-free.  
The top plane located near the headbox reveals the largest divergence levels and the bottom plane the smallest levels. 
Similar to Figs.~\ref{fig:v_x} and~\ref{fig:v_mag}, the discrepancy between the interpolated, time-averaged CFD results and the surrogate model predictions remains small. At the same time, the results illustrate that the modeling of the time-averaged propulsion influences by means of learned time-averaged velocity fields pursued here does not strictly enforce a divergence-free velocity field, since the target velocities are not continuity-compatible.

\subsection{Verification of Propulsion Model Integration into CFD}\label{sec:verification_SM_CFD}
To verify the proposed forcing-based propulsion modeling, numerical experiments are conducted on a specific SOV geometry. The cruising speed and geometric parameters employed for these studies, which are also part of the test data, are summarized in Table~\ref{tab:condition_values_design0}. We refer to this geometry as Design 0. All simulations discussed herein refer to this geometry, unless otherwise specified.

\begin{table}[ht]
\centering
\caption{Conditioning parameters used for the test case.}
\label{tab:condition_values_design0}
\begin{tabular}{lll}
\hline
\textbf{Parameter} & \textbf{Unit} & \textbf{Value} \\
\hline
Ship Speed         & m/s    & 6.73   \\
Stern Deadrise     & deg    & 0.0    \\
Stern Inclination  & deg    & 0.4    \\
SAY1               & deg    & 1.88   \\
SAY2               & deg    & 5.29   \\
HB\_Length         & -      & 1.0    \\
PhiX               & deg    & 0.0    \\
HBAS               & deg    & 11.95  \\
HBH                & m      & 0.42   \\
HBLB               & m      & 3.8    \\
HBLF               & m      & 2.12   \\
\hline
\end{tabular}
\end{table} 

\begin{figure}[ht!]  
    \centering
    \begin{subfigure}[b]{0.49\linewidth}
        \centering
        \includegraphics[width=\linewidth]{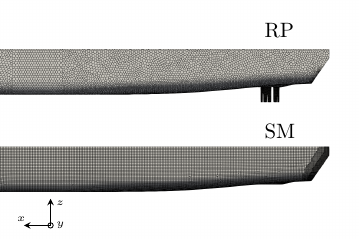}
        \caption{Side view of the hull.}
        \label{fig:grid_comp:side_view}
    \end{subfigure}
    \hfill
    \begin{subfigure}[b]{0.49\linewidth}
        \centering
        \includegraphics[width=\linewidth]{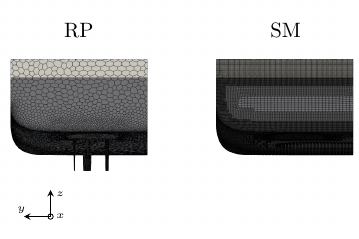}
        \caption{Transom view of the hull.}
        \label{fig:grid_comp:transom_view}
    \end{subfigure}
    \caption{Spatial discretization of ship hull for the RP and SM simulations.}
    \label{fig:grid_comparison}
\end{figure}

The initial setting refers to a simulation, in which both the ship and propeller geometries are resolved in the computational domain. 
We refer to this as the Resolved-Propeller (RP) simulation. This employs a numerical grid as described in Section~\ref{comp_domain_num_mesh}. Subsequently, the numerical grid resolving the propeller is substituted by the surrogate model, the remaining grid is refined and a set of pseudo-transient simulations are realized using a standard, pressure-based Finite Volume procedure  
\cite{rung2009} for different values of the non-dimensional forcing coefficient $\hat{\alpha}$. %
We refer to these as the Surrogate-Model (SM) simulations. The refined grid employed by the SM simulations involve approximately 1.6 million cells and targets to accurately capture near wall flow patterns in the vicinity of the ship (design). The different spatial discretizations between the RP and SM simulations are shown in Fig.~\ref{fig:grid_comparison}.

\begin{figure}[ht!]  
    \centering
    \begin{subfigure}[b]{0.43\linewidth}
        \centering
        \includegraphics[width=\linewidth]{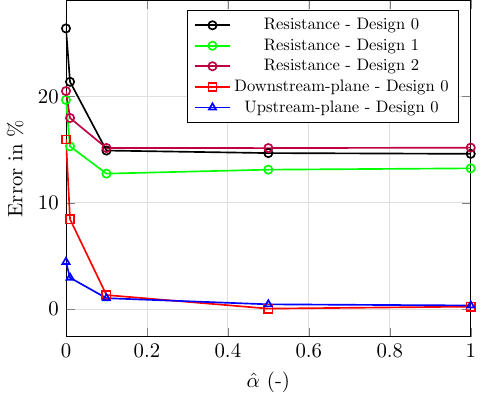}
        \caption{Resistance is shown in black, green and purple lines, the downstream-plane area-averaged $x$-velocity component in red line and the upstream-plane area-averaged $x$-velocity component in blue line.}
        \label{fig:error_CFD_RP_SM}
    \end{subfigure}
    \hfill
    \begin{subfigure}[b]{0.48\linewidth}
        \centering
        \includegraphics[width=\linewidth]{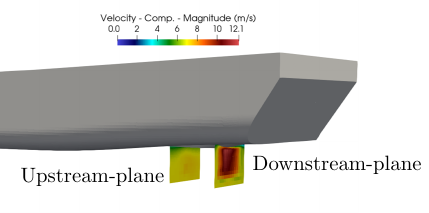}
        \caption{Sketch of the position and size of the planes used to compute the velocity metrics. Planes colored by computed velocity magnitude using the implicit forcing approach. Ship geometry corresponds to Design 0.}
        \label{fig:planes_sketch}
    \end{subfigure}
    \caption{Verification study for the propulsion model integration into CFD. \underline{Left}: Error in percentage of the SM-predicted metrics for different values of $\hat{\alpha}$ from the RP-predicted metrics.}
    \label{fig:error_CFD_RP_SM_overall}
\end{figure}

Both simulation setups measure the total resistance of the hull, i.e., the total force in the main flow direction ($x$-direction), supplemented by area-averaged velocity components $v_x$ extracted at two planes located slightly upstream and downstream of the propeller or surrogate model. Both planes are perpendicular to the $x$-direction and defined at a distance from the center of the VSP of approximately $\pm 0.75 \, D_\mathrm{VSP}$.
The depth and width of the planes are approximately, $L_\mathrm{VSP}$ and $1.37\,D_\mathrm{VSP}$, respectively. A sketch of the planes used to compute the area-averaged velocity metrics is shown in  Fig.~\ref{fig:planes_sketch}.

Figure~\ref{fig:error_CFD_RP_SM} shows the percentage error of the SM results relative to the RP results, which in this context we assume they refer to the "ground truth". The error is computed as $|\left(J^\mathrm{SM} - J^\mathrm{RP}\right)/J^\mathrm{RP}|\cdot 100$, where $J$ refers to one of the metrics described above. For small forcing coefficients, all metrics result in a significant error relative to the predictions of the RP simulation. Additionally, note that $\hat{\alpha} = 0$ corresponds to a simulation where the propeller influence is completely neglected. The error stagnates after $\hat{\alpha}= 0.1$, with further increases on the forcing coefficient resulting in minor changes on the predicted metrics. It is shown that as compared to a simulation in which the propeller is completely neglected ($\hat{\alpha} = 0$), the proposed solution forcing approach based on the ML-based surrogate model is able to produce significantly improved predictions, at a trivial (as compared to the RP simulation) computational cost. Specifically, an RP-based simulation requires on average 527 CPUh, while an SM-based simulation requires 175 CPUh for the same discretization of the ship’s geometry and the flow domain outside the propeller region. This corresponds to a $67$\% reduction in computational cost. However, this doesn't account for the complete training phase which amounts to approximately 95k CPUh, considering the simulation of the sample, the extraction process and the training. The break-even point is thus estimated at 270 SM simulations. This number can be rapidly met on an industrial level of application, especially when a more heuristic preliminary shape optimization strategy, e.g., based on evolutionary algorithms, is taken into account.

\begin{figure}[ht!]
    \centering
    \includegraphics[width=0.5\linewidth]{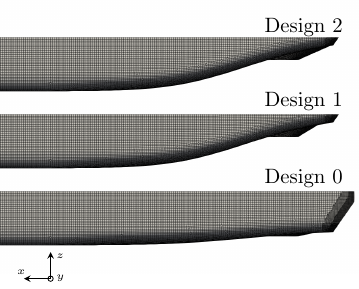}
    \caption{Comparative side view of the three designs used for the parametric $\hat{\alpha}$ study. Spatial discretization for SM simulations.}
    \label{fig:design012_side_view}
\end{figure}

 The error converges to approximately zero for the velocity-based metrics while it stagnates at approximately 14.5$\%$ for resistance. To further investigate the influence of the user-defined parameter $\hat{\alpha}$ on the prediction of the objective functional, we conducted the same parametric study on two more designs, namely Design 1 and Design 2. 
Both designs can be represented by the parametric surrogate model and, compared to Design 0, exhibit significant differences in the stern section of the ship, as shown in 
Fig.~\ref{fig:design012_side_view}. Figure~\ref{fig:error_CFD_RP_SM} shows that for all geometries, the value of resistance stagnates at approximately 15\% for any values of $\hat{\alpha}>0.1$.


While the magnitude of the latter is substantial, we would like to argue that it does not necessarily indicate a shortcoming of the surrogate model. As already discussed, the RP and SM simulations deliberately employ different grid discretizations. A comparatively coarse grid with approximately 0.9 million cells is used to discretize the domain in the RP case, excluding the mesh of the propeller. This allows us to reduce the overall offline training effort without compromising the accuracy of the propulsion data, since the propeller mesh is resolved using approximately 1.3 million cells, cf. Section~\ref{comp_domain_num_mesh}. A significantly finer grid with approximately 1.6 million cells is used for the SM case. The finer discretization is particularly better resolved in the near-wall region and along the walls, in view of the subsequent optimization studies, which require an accurate estimation of velocity gradients near the walls to compute the sensitivities. 

To investigate the influence of the mesh in the prediction of resistance, we performed one additional simulation using the discretization of the RP simulation without resolving the propeller. We then compared the resistance with the one predicted by the SM case for $\hat{\alpha}=0$. These two ''towing'' cases are equivalent and the resistance-prediction differences should reduce to the spatial discretization. The relative difference, using the same formula as the one used for the error computation, is found to be 14.6\% for the two towing cases. Therefore, we argue that the error of 14.5\% between the RP and the SM cases (for $\hat{\alpha} >0.1$) shown in Fig.~\ref{fig:error_CFD_RP_SM}, is largely due to the difference in discretization strategies and not due to the the surrogate model. Furthermore, based on a grid independence study conducted on the SM-based configuration it was found, that the employed discretization is sufficient to predict the objective without grid influences. Specifically, coarsening the employed grid to approximately 610k CVs results in an increase in the objective of about 10\%, i.e., much closer to the RP simulation, while a finer discretization of approximately 3.8 million CVs results in a decrease in the objective of less than 2\%. %

\begin{figure}[ht!]
    \centering
    \begin{subfigure}[b]{0.45\linewidth}
        \centering
        \includegraphics[width=\linewidth]{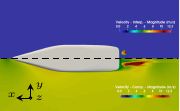}
        \caption{Contour plots at z = 0.1m below the mounting point of the VSP.}
        \label{fig:interpolation_verification_0.1}
    \end{subfigure}
    \hfill
    \begin{subfigure}[b]{0.45\linewidth}
        \centering
        \includegraphics[width=\linewidth]{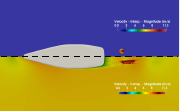}
        \caption{Contour plots at z = 0.9m below the mounting point of the VSP.}
        \label{fig:interpolation_verification_0.9}
    \end{subfigure}

    \caption{Velocity magnitude contours at 0.1m (left) and 0.9m (right) below the mounting point of the VSP. The dashed line denotes the symmetry plane. The top part is colored by the interpolated velocity magnitude coming from the surrogate model and the bottom part by the computed velocity magnitude as a result of the implicit forcing approach.}
    \label{fig:interpolation_verification}
\end{figure}

Figure~\ref{fig:interpolation_verification} shows the interpolated field (in the CFD mesh) and the computed velocity (in the CFD mesh) based on the employed solution forcing approach for $\hat{\alpha} = 0.5$ at a position of 0.1$\,$m (Fig.~\ref{fig:interpolation_verification_0.1}) and 0.9$\,$m (Fig.~\ref{fig:interpolation_verification_0.9}) below the mounting point of the VSP. As shown, all key features of the surrogate model-based flow prediction are sustained during the computation process while also being able to naturally blend the surrounding flow with the forced solution.

\section{Shape Optimization Method}
In the following, we describe the surrogate model shape optimization procedure considered in this work based on the employed adjoint system, the mesh deformation approach, and the handling of the additional geometric constraint. Finally an optimization algorithm for the implementation of the aforementioned is presented.%
\subsection{Adjoint System}
We consider a continuous adjoint, gradient-based and parameter-free shape optimization procedure. The objective function to be minimized is the total resistance of the underwater hull, viz.,
\begin{equation}
 J = \int_{\Gamma_\mathrm{O}} \left[p^\mathrm{eff} \delta_{ij} - 2\mu^\mathrm{eff}S_{ij}\right]n_j r_i \, \mathrm{d}\Gamma,   \label{eq:objective}
\end{equation}
where $\Gamma_\mathrm{O}$ denotes the boundary of the underwater hull, $n_j$ denotes the normal to the boundary and $r_i$ refers to the main flow spatial direction, in our case the $x$-direction. Note that when the surrogate model is considered, the primal solution is only solved using an unsteady setting to stabilize the solver, while in fact the solution is time-independent, due to the time-averaged velocity field coming from the surrogate model. Therefore, any temporal derivatives can be neglected. Furthermore, if the \textit{frozen turbulence} assumption, where variations in the turbulent viscosity are neglected, is applied in deriving the dual system for Eqs.~(\ref{eq:contrinuity}, \ref{eq:momentum}), the adjoint field equations take the form
\begin{align}
    \hat{R}^\mathrm{p} &= -\frac{\partial \hat{v}_i}{\partial x_i} = 0\label{eq:adj_continuity}\\
    \hat{R}^\mathrm{v}_i &= \rho \hat{v}_j\frac{\partial v_j}{\partial x_i} -\rho v_j \frac{\partial \hat{v}_i}{\partial x_j} - \frac{\partial}{\partial x_j}\left(2\mu^\mathrm{eff}\hat{S}_{ij} - \hat{p}\delta_{ij}\right) = 0,\label{eq:adj_momentum}
\end{align}

where $\hat{v}_i$ and $\hat{p}$ denote the adjoint velocity components and adjoint pressure, respectively. The components of the adjoint strain-rate tensor are denoted by $\hat{S}_{ij}$. The system of equations is closed by the following boundary conditions for adjoint velocity and adjoint pressure,
\begin{align}
    \hat{v}_i &= - r_i,  &&\frac{\partial \hat{p}}{\partial x_i}n_i = 0 \qquad \text{on the ship hull, i.e. $\Gamma_\mathrm{O}$} \nonumber \\
    \hat{v}_i &= 0,  &&\frac{\partial \hat{p}}{\partial x_i}n_i = 0  \qquad  \text{on the inlet,} \nonumber \\
    \frac{\partial \hat{v}_i}{\partial x_k}n_k &= 0,  &&\hat{p} = \rho v_k n_k \hat{v}_i n_i  \quad \text{on the outlet} \quad \text{and} \nonumber \\
    \hat{v}_i n_i & = 0,  && \frac{\partial \hat{v}_i}{\partial x_j}n_jt_i = 0, \quad \frac{\partial \hat{p}}{\partial x_i}n_i = 0  \quad \text{on symmetry plane and slip walls,} \label{eq:adj_BCs}
\end{align}
where $t_i$ denotes the tangential vector to the surface. A detailed derivation of the adjoint field equations and the corresponding boundary conditions for the employed objective can be found in \cite{5_kuhl2022adjoint}.

The system of adjoint equations is discretized using a similar FVM procedure as the one used for the primal.  The adjoint pressure-velocity coupling is resolved by a method similar to the primal SIMPLE algorithm. Note that in adjoint systems, the information transfer is reversed w.r.t. the primal. To this end, the convective term is discretized by a downwind analogy to the corresponding primal discretization scheme.

After the solution of the primal and adjoint
system of equations, the adjoint sensitivity can be computed as
\begin{equation}
    s = -\int_{\Gamma_\mathrm{O}} \mu^\mathrm{eff} \frac{\partial v_i}{\partial x_j}n_j \frac{\partial \hat{v}_i}{\partial x_k}n_k \, \mathrm{d}\Gamma. \label{eq:adj_sensitivity}
\end{equation}
The detailed derivation of Eq.~(\ref{eq:adj_sensitivity}) can be found, for instance in \cite{othmer2008,kuhl2019}. Based on this expression, we can also define a local sensitivity that can be computed in each discretized boundary face of $\Gamma_\mathrm{O}$ as,
\begin{equation}
    s^\mathrm{loc} = \mu^\mathrm{eff} \frac{\partial v_i}{\partial x_j}n_j \frac{\partial \hat{v}_i}{\partial x_k}n_k \quad \text{on  $\Gamma_\mathrm{O}$}.\label{eq:adj_local_sensitivity}
\end{equation}
%
%
Once the physical sensitivity is computed, the process continues by computing an admissible deformation field for every computational node of the mesh. A variety of different such approaches can be found in \cite{radtke2023}. In this work, we employ the the Steklov-Poincar\'e method to obtain the deformation field. The latter as well as the approach to conserve the submerged volume of the hull are outlined in \cite{5_kuhl2022adjoint}.

The shape optimization algorithm employed in this work is summarized in Algorithm~\ref{alg:optimization_algorithm}. In lines 1-4, the algorithm proceeds to employ the pre-trained surrogate model given the conditioning parameters, geometric and flow conditions, that relate to the optimization case. Note that the surrogate model is only employed once, outside of the main optimization loop. This implies that any changes in the geometry of the ship are not taken into account for the subsequent optimization steps. The main optimization loop (lines 5-24) incorporates an Armijo backtracking line search, which is summarized by the loop in lines 6-15. Specifically to that, the initial step size $t^0$ is implicitly determined on the first optimization iteration by a maximum displacement approach, in which the maximum displacement $d^{\mathrm{max}}$ is usually set to $L/100$ or $L/1000$, where $L$ corresponds to a characteristic length of the geometry. The algorithm then checks whether the condition in line 9 is fulfilled based on a prescribed value of $c \in (0,1)$ and if not, it reduces the step size by $\tau \in (0,1)$. We consider the optimization converged if the relative decrease of the objective function between two consecutive steps is below $\epsilon^\mathrm{conv}$. 

Due to the parameter-free approach employed by this work, the geometry does not require re-meshing and the initial topology is sustained. This feature of the method enables 
\begin{enumerate}
    \item the consistent application of the surrogate model-based velocity field without the need of re-interpolating for each design candidate
    \item restarting the primal and adjoint simulations based on the solutions of the previous iteration.
\end{enumerate}    
The latter results in a significant speed-up of the whole procedure since the required (pseudo) time steps for convergence of the primal problem are considerably reduced in comparison to the initial shape.
\begin{algorithm}[ht!]
\caption{Algorithm for adjoint-assisted, gradient-based, parameter-free shape optimization using a propulsion surrogate model. Here, $dJ$ refers to the gradient of the objective functional w.r.t. the shape.}
\label{alg:optimization_algorithm}
\begin{algorithmic}[1]
	\Require \raggedright initial shape, $d^{\mathrm{max}}$, $\epsilon^\mathrm{conv} > 0$, $\hat{\alpha}\geq0$, $c, \tau \in (0,1)$, $N^\mathrm{max}$, $L^\mathrm{max}$
    \State Identify conditioning parameters based on case, cf. Table~\ref{tab:condition_ranges}.
    \State Call surrogate model to acquire velocity field in the local meta-grid.
    \State Translate and rotate the meta-grid to align with the mounting point of the propeller in the CFD mesh.
    \State Interpolate the velocity field to the CFD mesh based on Eq.~(\ref{eq:nearest_neigh}).
	\For{$k=0,1,...$}
        \For{$l=0,1,...$}
		    \State Solve the  primal problem using solution forcing in the momentum equations, cf. Eq.~(\ref{eq:forcing_mod}).
            \State Evaluate objective function $J^{k}$, cf. Eq.~(\ref{eq:objective}).
            \If{($k>0$ and $J^{k} \le J^{k-1} + c \cdot t^{k-1} \cdot dJ^{k-1}$) or ($l>L^\mathrm{max}$) or ($k=0$)}
               \State \textbf{break}
            \Else
               \State $t^k \leftarrow \tau t^{k} $
               \State Reverse mesh to level $k-1$ and re-update based on new step size.
            \EndIf 
        \EndFor
        \If{($k>0$ and $\left(|J^k - J^{k-1}|/J^{k-1}\right)\cdot 100 \leq \epsilon^\mathrm{conv}$) or $(k > N^\mathrm{max})$}
           \State \textbf{break}
        \EndIf
        \State Solve the adjoint problem, cf. Eq.~(\ref{eq:adj_continuity}, \ref{eq:adj_momentum}).%
        \State Evaluate local sensitivity, Eq.~(\ref{eq:adj_local_sensitivity}).
        \State Compute descent direction based on Steklov-Poincar\'e method and volume conservation constraint, see e.g., \cite{5_kuhl2022adjoint}. %
        \State Update mesh. 
	\EndFor
    \end{algorithmic}
\end{algorithm}
%

\FloatBarrier
\section{Optimization Application}\label{sec:opt_app}
\begin{figure}[ht!]
    \centering
    \begin{subfigure}[b]{0.52\linewidth}
        \centering
        \includegraphics[width=\linewidth]{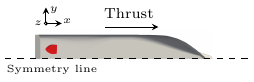}
        \caption{Top view of the geometry.}
        \label{fig:top_view}
    \end{subfigure} 
    \hfill
    \begin{subfigure}[b]{0.41\linewidth}
        \centering
        \includegraphics[width=\linewidth]{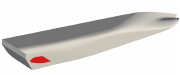}
        \caption{Perspective view of the geometry.}
        \label{fig:perspective_view}
    \end{subfigure}
    \hfill
    \begin{subfigure}[b]{0.52\linewidth}
        \centering
        \includegraphics[width=\linewidth]{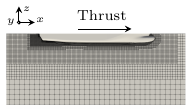}
        \caption{Side view of the discretized computational domain near the ship.}
        \label{fig:grid}
    \end{subfigure}
    \caption{Initial shape and discretized computational domain. Areas of the hull that are free for design are displayed in grey while those that are fixed to their initial position are displayed with red.}
    \label{fig:geometry_sketch}
\end{figure}

The proposed surrogate model-based optimization method is applied to optimize the shape of an SOV. The initial hull geometry of the vessel corresponds to the test case investigated in Section \ref{sec:verification_SM_CFD}. It is described by the parameters in Table~\ref{tab:condition_values_design0} and illustrated in Figs.~\ref{fig:top_view} and \ref{fig:perspective_view}.
Note that the headbox of the SOV, displayed in red, is not a design surface,
to ensure that a plane surface is sustained during the optimization on which the VSP propeller can be properly mounted. Simulations are performed for a half model at a ship speed of $V_s=6.73\, \mathrm{m/s}$, resulting in a $\mathrm{Re}$-number of approximately $6.35 \cdot 10^8$.   The  computational domain is discretized with approximately 1.6 million locally-refined hexahedral control volumes. Figure~\ref{fig:grid} shows the computational grid used near the hull of the ship. As shown there, the grid is progressively refined towards the ship so that to accurately predict the boundary layer along the hull. However, there is no extra artificial refinement in the area swept by the propeller. 
A $k-\epsilon$ Boussinesq-viscosity model is used to model turbulence. A QUICK scheme is used to discretize all convective terms while pseudo-time stepping is performed for primal convergence using an implicit Euler temporal discretization with a time step of $\Delta t \cdot V_s/L_{OA} =0.365 \cdot 10^{-3}$. All simulations herein are realized using the CFD solver FreSCo$^+$ \cite{rung2009}.

In the optimization process, the initial step size is determined using a maximum displacement approach that constricts the maximum local displacement on a computational face to $d^\mathrm{max} = \mathrm{LOA}/1000$. Convergence is monitored by $\epsilon^\mathrm{conv}=0.1$. This implies that the optimization is terminated when the objective for two subsequent shapes changes by less than 0.1\%. Optimization and Armijo thresholds are set to $N^\mathrm{max} = 50$ and $L^\mathrm{max} = 10$, respectively. Armijo constants are set to $c = 10^{-20}$ and $\tau = 0.5$, effectively checking only cases in which the objective function increases in two consecutive steps. If the objective function increases, the step size is halved and the simulation is repeated until the condition of decreasing objective is met. An increasing objective can happen due to the application of the volume constraint, which alters the deformation field in such a way that a descent direction is not always guaranteed. Furthermore, we would like to note that while $\tau = 0.5$ corresponds to a usual choice, $c= 10^{-20}$ is less common, with typical values ranging around $10^{-4}$. Due to the expensive nature of the full-scale CFD simulations realized in the iterative process of optimization, the strict selection of $c=10^{-20}$ is made to avoid potentially unnecessary Armijo iterations. 

We consider two optimization cases. The first corresponds to an optimization which neglects the propeller influence and is realized by setting $\hat{\alpha}=0$. We refer to this as the \textit{No-Propeller} (NP) case. The second run employs a forcing coefficient of $\hat{\alpha} = 0.5$, based on the results presented in Fig.~\ref{fig:error_CFD_RP_SM}. We refer to this as the \textit{Surrogate-Propeller} (SP) case. 
\subsection{Optimization results}
\begin{figure}[ht!]
    \centering
    \begin{subfigure}[b]{0.49\linewidth}
        \centering
        \includegraphics[trim={2cm 0 0 0}, clip, width=\linewidth]{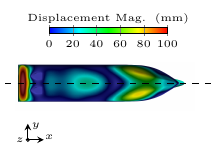}
        \caption{Top view of the displacement field for NP.}
        \label{fig:displ_NP_top}
    \end{subfigure} 
    \hfill
    \begin{subfigure}[b]{0.49\linewidth}
        \centering
        \includegraphics[trim={2cm 0.0cm 0.0cm 0.0cm}, clip, width=0.98\linewidth]{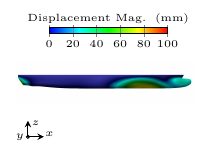}
        \caption{Side view of the displacement field for NP.}
        \label{fig:displ_NP_side}
    \end{subfigure}
    \hfill
    \begin{subfigure}[b]{0.49\linewidth}
        \centering
        \includegraphics[trim={2cm 0 0 0}, clip, width=\linewidth]{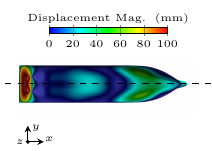}
        \caption{Top view of the displacement field for SP.}
        \label{fig:displ_FS_top}
    \end{subfigure} 
    \hfill
    \begin{subfigure}[b]{0.49\linewidth}
        \centering
        \includegraphics[trim={2cm 0.0cm 0.0cm 0.0cm}, clip, width=0.98\linewidth]{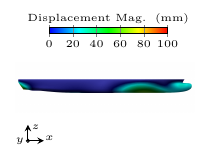}
        \caption{Side view of the displacement field for SP.}
        \label{fig:displ_FS_side}
    \end{subfigure}
    \hfill
    \caption{Displacement field computed on the hull for the first optimization iteration ($k=0$). Complete hull is used for visualization purposes. Dashed line indicates the symmetry plane.}
    \label{fig:displacement_01}
\end{figure}

Figure~\ref{fig:displacement_01} shows the computed (constrained) displacement field on the hull for the first optimization iteration for each case. Three characteristic areas can be identified as the most influential to the optimization process: the transom (leftmost part of the hull), the midship and the bow. The largest contribution is observed at the transom, regardless of the optimization case. Interestingly, when the surrogate model is considered, an additional region of high displacement appears near the propeller wake. Moreover, the displacement fields at the midship and bow also significantly differ in both their magnitude and pattern.

\begin{figure}[ht!]
    \centering
    \begin{subfigure}[b]{0.45\linewidth}
        \centering
        \includegraphics[width=\linewidth]{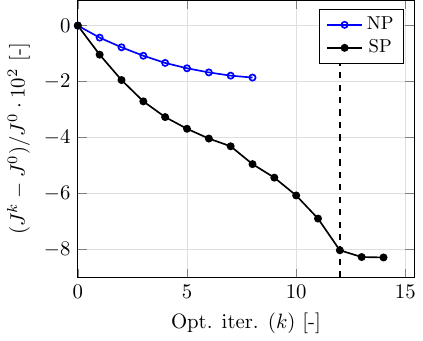}
        \caption{Relative decrease of resistance in \%.}
        \label{fig:objective_history}
    \end{subfigure} 
    \hfill
    \begin{subfigure}[b]{0.45\linewidth}
        \centering
        \includegraphics[width=\linewidth]{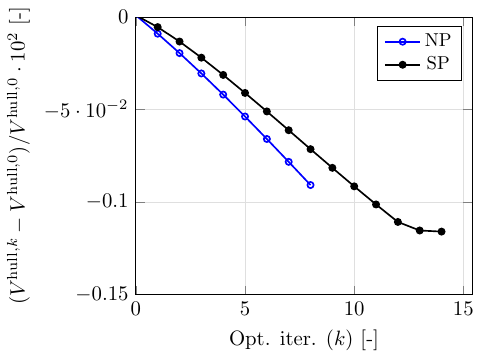}
        \caption{Relative change of displaced volume in \%.}
        \label{fig:volume_history}
    \end{subfigure}
    \caption{Optimization history. Vertical dashed lines highlight the optimization iteration in which the Armijo condition was for the first time not fulfilled.}
    \label{fig:optimization_history}
\end{figure}

Figure~\ref{fig:objective_history} shows the relative decrease of the objective function as compared to the initial shape during the optimization process. Both cases met the imposed convergence criterion with NP reaching a decrease of approximately 2\% while SP reaches a decrease of approximately 8.2\%. %
The SP case did not satisfy the Armijo condition at iteration 12 based on the initial step size. However, after a maximum of 4 Armijo iterations, a smaller step size able to satisfy the condition was successfully identified. Figure~\ref{fig:volume_history} shows the relative change of the displaced water volume (submerged volume of the hull) during the optimization. Both constrained optimization cases managed to sustain the initial volume with no changes larger than 0.15\% reported.

It is additionally important to note that in Fig.~\ref{fig:objective_history} we show the relative decrease of each case based on their corresponding primal. This means that while the SP case predicts the resistance with the use of the surrogate model, the NP prediction disregards this and thus $J^\mathrm{0,NP} \neq J^\mathrm{0,SP}$.

\begin{figure}[ht!]
    \centering
    \begin{subfigure}[b]{0.8\linewidth}
        \centering
        \includegraphics[width=\linewidth]{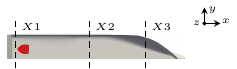}
        \caption{Geometric annotations on half-ship sketch for the qualitative position of the framelines.}
        \label{fig:geom_frame}
    \end{subfigure} 
    \hfill
    \begin{subfigure}[b]{0.32\linewidth}
        \centering
        \includegraphics[width=\linewidth]{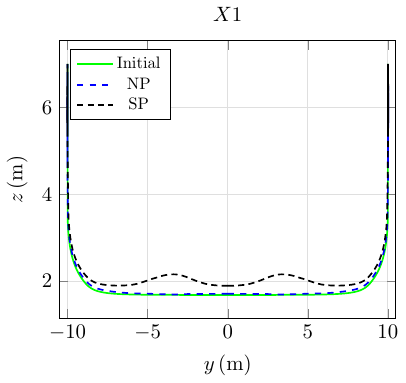}
        \caption{Framelines at $X1$.}
        \label{fig:frame_x1}
    \end{subfigure}
    \hfill
    \begin{subfigure}[b]{0.32\linewidth}
        \centering
        \includegraphics[width=\linewidth]{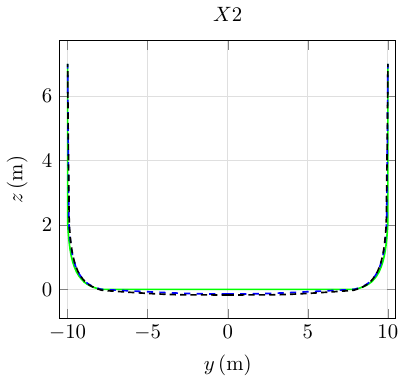}
        \caption{Framelines at $X2$.}
        \label{fig:frame_x2}
    \end{subfigure}
    \hfill
    \begin{subfigure}[b]{0.32\linewidth}
        \centering
        \includegraphics[width=\linewidth]{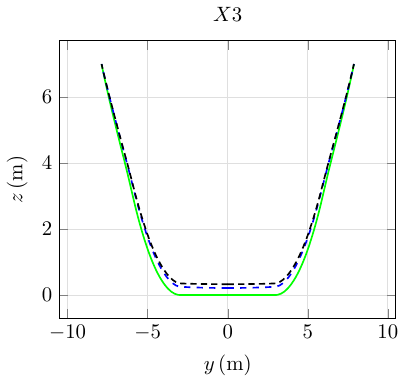}
        \caption{Framelines at $X3$.}
        \label{fig:frame_x3}
    \end{subfigure}
    \caption{Framelines at three characteristic positions of the ship along its length. Continuous (green) line corresponds to the initial shape, dashed (blue) line to the optimized shape of the NP case, densely dashed (black) line to the optimized shape of the SP case %
    Symmetry line at $y=0$.}
    \label{fig:framelines}
\end{figure}

In Fig.~\ref{fig:framelines} we show the framelines at three characteristic positions of the ship along its length. As expected by the predicted displacement field (Fig.~\ref{fig:displacement_01}), the most noticeable difference between the NP and SP cases are observed near the wake of the propeller. When a surrogate model is considered, the optimizer proceeds to create a "tunnel-like" segment behind the headbox allowing for a smoother flow transition towards the transom. Additionally, both cases resulted in a minimization of the wetted area near the $X3$ position. Finally, the changes at the midship region were trivial as compared to $X1$ and $X3$.

\subsection{Validation of the Optimization Studies}
The validation of the optimization studies aims to evaluate the relevance of considering propulsion effects in studies of hull-shape optimization. For this purpose, the optimal shapes of the two optimization studies with propulsion influences (SP) and without propulsion influences (NP) are analyzed. The analysis of both shapes is based on two corresponding primal flow simulations: a transient simulation with a geometrically resolved VSP, cf. Section~\ref{cfd_dataset}, and a steady-state simulation using the ML-based surrogate model of the VSP. Reported results are restricted to the objective functional, i.e., the hydrodynamic drag obtained from these four primary flow studies, and their comparison with the prognosis of the adjoint shape optimization.

Results obtained from the ML-based surrogate model simulations predict an \textbf{increase}  in drag by approximately $3$\% for the shape obtained from the NP-optimization, which clearly contradict the prognosis of the optimization study, and a \textbf{decrease} of about $8.2$\% for the SP-optimization. 
Finally, 
the optimized shapes are assessed using transient simulations in which the propeller is explicitly resolved.
The resolved-propeller simulation of the NP-optimized shape
predicted a $2.26\%$ \textbf{increase} in resistance 
and also clearly contradicts the optimization study. 
In contrast, the corresponding simulation for the SP-optimized shape 
predicted a $9.29\%$ \textbf{decrease}. 

Looking at the shape obtained from an optimization study that neglects propulsion influences, we note that the  drag for the (supposedly) optimized hull shape actually increased by approximately 2\%-3\% in both simulations. This result underscores the need to include propulsion influences in a hull-shape optimization process. On the other hand, both simulations confirm a significant decrease in resistance about 8\%-9 \% for the hull shape derived from the optimization with propulsion.
Although the relative values obtained from the resolved and ML-based propulsion simulations do not exactly agree,  
the overall trend is remarkably accurate considering the fundamentally different grids. This agreement reinforces the efficiency of the surrogate-model-based optimization approach, demonstrating that additional expensive high-fidelity, propeller-resolved CFD simulations are not required to confirm the optimization outcomes.

Furthermore, it is interesting to note the response of the optimized shapes to an update of surrogate model conditions. The results of a mesh-based re-evaluation of the geometric parameters of the optimized NP and SP shapes is shown in Table~\ref{tab:condition_values_change}. 
\begin{table}[ht]
\centering
\caption{Change of conditioning parameters from initial to optimized shape. Initial parameters correspond to those presented in Table~\ref{tab:condition_values_design0}.}
\label{tab:condition_values_change}
\begin{tabular}{lll}
\hline
\textbf{Parameter} & \textbf{Change - NP (\%)} & \textbf{Change  - SP (\%)} \\
\hline
Ship Speed ($x$-direction)        & -     & -   \\
Stern Deadrise     & -    & -   \\
Stern Inclination  & -    & -  \\
SAY1               & -8.60    &  59.05  \\
SAY2               & -8.08    & -8.07  \\
HB\_Length         & -      & -   \\
PhiX               & -      & -   \\
HBAS               & -0.20   &  2.01  \\
HBH                & 1.80      & 9.56   \\
HBLB               & 0.18      & 0.47  \\
HBLF               & -      & -   \\
\hline
\end{tabular}
\end{table}

Surrogate-based simulations with the updated geometric conditions for the optimized shapes showed the following:
\begin{itemize}
    \item For the NP-based optimized shape, the re-evaluation showed an \textbf{increase} in drag by approximately $0.03 \%$ as compared to the surrogate evaluation with frozen geometric parameters. 
    \item For the SP-based optimized shape, the re-evaluation showed a \textbf{decrease} in drag by approximately $0.11 \%$ as compared to the surrogate evaluation with frozen geometric parameters. 
\end{itemize}
These results show that the sensitivity of the objective to the surrogate input geometric parameters is relatively small. However, it must be noted that this does of course not guarantee that the same holds for the sensitivity of the gradient of the objective w.r.t. the shape. Therefore, a consistent surrogate integration in the optimization loop needs to be further investigated.

Finally, to reduce the complexity of the optimization problem, simplifying assumptions such as a single-phase flow and a fixed floating position are employed in this work. However, it is interesting to examine the performance of the optimized shape in more realistic operational conditions. 
To this end, the optimized shape of the SP optimization study is suitably remeshed to capture free surface phenomena and to allow for a two-phase flow simulation of a geometrically resolved VSP simulation. In addition, this simulation considers a free-floating vessel and a self-propulsion operation through a controller of the VSP rotational speed. The simulation reports 
a \textbf{decrease} in resistance of about 22.28\%.
In view of the self-propulsion condition, it is also interesting to note that the thrust deduction factor is \textbf{decreased} by approximately 31\% for SP-optimized shape, respectively. 
The latter is a naval architecture parameter representing the increase of resistance caused by the suction effect of the propeller and should ideally become small.
Therefore, it is shown that for the case of more realistic operational conditions, the optimal shape outperforms the initial one.
Finally, we would like to note that while this observation holds for the investigated application, it can not be generalized for all vessels of this type and all operating conditions. To this end, a shape optimization process considering the more involved realistic operational conditions in both primal and adjoint should be considered.


\FloatBarrier
\section{Conclusion}

In this study, a parametric surrogate model for the Voith Schneider Propeller (VSP) was developed to take into account the influences of propeller-hull interaction within a hull-shape optimization process, that aims to minimize the hydrodynamic resistance of a 
Service Operation Vessel (SOV) equipped with a five-bladed Voith–Schneider Propeller (VSP) in full-scale. 

By using the data-driven surrogate model in an adjoint-based shape optimization framework, the optimization procedure can correctly mimic the propeller-hull interaction and thus gains considerable accuracy compared to an optimization without considering the surrogate model.

During the offline phase, 180 unsteady Reynolds-averaged Navier–Stokes simulations were performed and the velocity fields for each training sample were time-averaged and mapped from the unstructured CFD grids onto a structured meta-grid.

Subsequently, a conditional variational autoencoder (CVAE) was developed to model the mapped three-dimensional velocity fields based on 10 geometric conditions and the cruising velocity. Residual blocks and self-attention mechanisms were incorporated into the encoder and decoder to enhance the model’s ability to capture strong local gradients. The validation of the surrogate model for unseen data reveals an excellent agreement with the time-averaged full-order model results. 
It was shown that the maximum relative percentage error across all slices of the trained model does not exceed 1.6\%
The comparison shown also demonstrates that the time-averaged velocity fields cannot be continuity-compatible because of the temporary overlap of the fields by the propeller blades. This must be taken into account during further data processing.

In the second phase, the surrogate model was used to support steady primal 
simulations, which are required to optimize the hull shape. 
To this end, the velocity field generated by the surrogate model was interpolated in a mesh of finer resolution in the vicinity of the hull that does not geometrically resolve the propeller. The machine learning-based velocity field was implicitly incorporated into the solution of the discrete momentum equations. This implicit forcing is based on local weights and yields very good agreement with the learned velocities without enforcing 100\% agreement, which is undesirable due to the aforementioned continuity issues.
Specifically, it was found that already for a  non-dimensional forcing coefficient slightly above $\hat{\alpha} = 0.1$ the surrogate model-based approach was able to %
reproduce several metrics of interest for the unsteady flow around the initial vessel under investigation, with differences being attributable to discrepancies in the spatial discretization.

Finally, the proposed methodology was integrated into a CAD-free shape optimization process consisting of sequences of primal and adjoint simulations and applied to the SOV geometry. Our results showed that: 
\begin{enumerate}
    \item omitting the propulsion system during optimization can be detrimental. While the optimization run reported a 2\% decrease of the resistance when neglecting the propulsion system, subsequent CFD simulations including the propulsor showed that the optimized ship's resistance was actually increased by approximately 3\% as compared to the initial shape.
    \item the surrogate model was able to identify hull geometries that lead to a decrease of the ship's resistance by more than 8\%. This result was also validated by a propeller-resolved simulation.
\end{enumerate}

These results highlight the importance of considering the propeller–hull interaction in adjoint-based hull optimizations and demonstrate the effectiveness of coupling data-driven surrogate models with physics-based CFD solvers.

Limitations of the displayed investigations include the “frozen” treatment of the surrogate model during optimization and the single-phase simulation approach used for the CFD and optimization simulations shown in this work. 

We are currently working towards updating the geometric conditions that serve as input to the surrogate model at each design candidate during an optimization run, i.e., including lines 1-4 of Algorithm ~\ref{alg:optimization_algorithm} inside the optimization loop (lines 5-23). We also target to extend the proposed methodology to two-phase flows and self-propulsion settings that allow dynamic movement of the vessel. Finally, we are working towards enhancing the consistency of the adjoint solver by explicitly differentiating through the discrete forcing method.


\bibliographystyle{elsarticle-num}

\section*{Acknowledgements}

The authors would like to thank Henning Schwarz for his valuable discussions and contributions to the development of the surrogate model, as well as David Bendl for his assistance in preparing the geometries.

M.A.M. acknowledges the support of the Marie Skłodowska-Curie Actions under the HORIZON-MSCA-2021-DN-01 programme (Grant No. 101072851), through the MFLOPS (Multiphase Flow Optimisation Strategies with Industrial Applications) project.
G.B., A.H., T.T.N., M.P.  and T.R. acknowledge support from the "Propulsion Optimization of Ships and Appendages" (ProSA) research project funded by the German Federal Ministry for Economic Affairs and  Energy (BMWE; Grant No. 03SX599C).

\section*{Author contributions}
\textbf{Moloud A. M.:} Writing – original draft, Methodology, Software, Investigation, Formal analysis, Validation;
\textbf{Georgios B.:} Writing – original draft, Writing – review \& editing, Methodology, Software, Investigation, Formal analysis, Validation;
\textbf{Thanh T. N.:} Data curation, Investigation, Writing – review \& editing, Validation;
\textbf{Ahmed H.:} Software, Investigation;
\textbf{Michael P.:} Supervision, Funding acquisition ;
\textbf{Thomas R.:} Writing – review \& editing, Supervision, Project administration, Funding acquisition.

\bibliography{references}
\end{document}